\documentclass[journal]{IEEEtran}

\usepackage{amsmath}
\usepackage{amssymb}
\usepackage{amsfonts}
\usepackage{graphicx}
\usepackage{cite}
\usepackage{algorithm}
\usepackage{algorithmic}
\usepackage{booktabs}
\usepackage{multirow}
\usepackage{url}
\usepackage{placeins}
\usepackage[caption=false,font=footnotesize]{subfig}
\usepackage{dblfloatfix}

\usepackage{acronym}
\usepackage{etoolbox}

\input{Acronymes}

\begin{document}

\title{Domain-Aware Hierarchical Federated Learning for UAV Swarms Across Heterogeneous Environments}

%%%%%%%%%%%%%%%%%%%%%%%%%%%
\author{Taha Mahmood, Mudassar Shakeel, Ahmed~Naeem, El Mehdi Amhoud~\IEEEmembership{Senior Member,~IEEE}, Rizwan Ahmad~\IEEEmembership{Member,~IEEE}, Salman Ghafoor \\This work has been submitted to the IEEE for possible publication. Copyright may be transferred without notice, after which this version may no longer be accessible.

\thanks{T. Mahmood, M. Shakeel, A. Naeem, R. Ahmad, and S. Ghafoor are with the School of Electrical Engineering and Computer Science, National University of Sciences and Technology, 44000, Islamabad, Pakistan. (email: \{tmahmood.bee23seecs, mshakeel.bee23seecs,
ahmed.naeem, rizwan.ahmad, salman.ghafoor\}@seecs.edu.pk). E. M. Amhoud is with the College of Computing, University Mohammed VI Polytechnic, Benguerir 43150, Morocco (email: elmehdi.amhoud@um6p.ma).}
}

\maketitle
\begin{abstract}
Hierarchical federated learning (HFL) enables communication-efficient training in unmanned aerial vehicle (UAV) swarms, but performance degrades under changing and heterogeneous environments. This paper proposes domain-aware HFL (DA-HFL), integrating mobility-informed clustering, cluster-level aggregation, cloud-based Fisher-weighted merging, a domain-adversarial shared backbone, environment-specific heads, and gated inference with local adaptation. A mission-weighted optimization formulation incorporates energy, latency, and cluster-head eligibility constraints. Theoretical analysis establishes resource feasibility, a conditional backbone-stationarity bound, the quadratic optimality of Fisher-weighted merging, and communication and computation complexity bounds. Simulations across four heterogeneous environments demonstrate faster convergence, improved accuracy during environmental transitions, and more stable, purer clusters than the evaluated baselines, with modest energy overhead over resource-aware HFL. Ablation studies validate the contributions of the proposed components.
\end{abstract}

\section{Introduction}

\Ac{6G} wireless networks are envisioned to provide intelligent,
distributed, and autonomous connectivity through low-latency
communication, edge intelligence, and \ac{ISAC} \cite{saad2020vision}. \Ac{UAV}
swarms are an important use case because of their
flexible deployment, and distributed sensing capabilities.
They support applications such as disaster response, surveillance, and
environmental monitoring where \acp{UAV}
must process sensor data and make decisions during flight.
Transmitting these data to a remote server introduces communication
overhead and latency \cite{liu2020client} over intermittent or
resource-constrained links. Enabling learning and inference
at the network edge is therefore important for
\ac{UAV} operations.

\Ac{FL} allows \acp{UAV} to retain raw data locally and exchange
model parameters or updates with an aggregation server
\cite{mcmahan2017communication}. \Ac{HFL} introduces intermediate
cluster-level aggregation between \acp{UAV} and the cloud to
reduce direct server communication and support larger networks
\cite{liu2020client}. Resource-aware approaches use energy,
mobility, computation, and link conditions to guide clustering
and dynamic \ac{CH} selection
\cite{he2023threestage,wang2024uav,shen2023joint}.
Asynchronous aggregation further reduces synchronization delays
caused by straggling participants
\cite{10452821,wu2020safa}.

However, resource-aware organization alone does not address changes
in the data distribution observed by a moving \ac{UAV}. Statistical
heterogeneity across clients can degrade the performance of a common
federated model \cite{10065509}, while feature-distribution
shifts further motivate domain-specific treatment
\cite{li2021fedbn}. During a mission, a \ac{UAV} may move across
urban, forest, rural, and lake regions, each associated with a
different data distribution. Consequently, a model suited to one
environment may become less effective after an environmental
transition. Existing resource-aware clustering mechanisms
\cite{he2023threestage,wang2024uav,shen2023joint} primarily organize
\acp{UAV} according to communication, mobility, energy, and
computation conditions, and therefore do not explicitly determine
when an environmental-domain change requires model reassignment.
Moreover, aggregating heterogeneous updates into a single model can
dilute domain-specific knowledge, whereas a static model assignment
may retain a predictor associated with a previously occupied
environment.

\subsection{Related Work}

Resource-aware \ac{HFL} for \acp{UAV} addresses limited battery
capacity, heterogeneous computation, and time-varying wireless links.
Existing approaches combine energy-aware, mobility-aware, and channel-aware clustering with dynamic \ac{CH} selection to improve aggregation
efficiency \cite{he2023threestage,wang2024uav,shen2023joint}.
Asynchronous and event-driven frameworks further reduce the need to
wait for all participants before aggregation
\cite{10452821,wu2020safa}. However, these mechanisms
primarily characterize resource and communication conditions rather
than changes in the sensed data domain. Thus, resource-feasible
clusters may still require model reassignment when \acp{UAV} move
between environments.

The statistical heterogeneity in \ac{FL} can impair global training
\cite{10065509}, motivating clustered and optimization-based
solutions. Clustered \ac{FL} groups clients according to learning
updates and maintains specialized models
\cite{sattler2021clustered}, while FedProx uses a proximal penalty
to improve training under heterogeneous data
\cite{li2020federated}. SCAFFOLD uses control variates to reduce
client drift and provides convergence guarantees under its stated
assumptions \cite{karimireddy2020scaffold}. These methods address
heterogeneous optimization, but do not explicitly determine when a
mobile \ac{UAV} should switch between environment-specific models.

Domain adaptation and personalized \ac{FL} provide complementary
mechanisms for heterogeneous observations. \Ac{FedBN} retains
batch-normalization components locally to accommodate feature shift
across clients \cite{li2021fedbn}, while \Ac{DANN} uses a domain
discriminator and gradient reversal to learn less domain-dependent
representations \cite{ganin2016domain}. Domain-generalization
methods similarly seek transferable representations
\cite{wang2022generalizing}, whereas Ditto and Per-FedAvg learn
client-specific models or adaptable initializations
\cite{li2021ditto,fallah2020personalized}. These approaches support
invariance or local specialization, but do not directly connect
environmental transitions with hierarchical aggregation and model
assignment.

\Ac{MoE} architectures maintain specialized models and use a gate to
select or combine their outputs \cite{11303974},
with adaptive expert selection also explored in federated settings
\cite{yi2026pfedmoe}. For mobile \ac{UAV} hierarchies,
however, expert selection must also account for model availability,
communication, and environmental changes. Parameter-importance
methods provide another mechanism for aggregation: Fisher-weighted
merging assigns coordinate-dependent weights to model parameters
\cite{matena2022merging}, while Fisher-based regularization preserves
previously learned knowledge \cite{kirkpatrick2017overcoming}.
These methods motivate importance-aware aggregation, but do not
specify how environment-specific heads should be grouped, propagated
across hierarchy levels, or selected after a \ac{UAV} changes its
environment.
The remaining requirement is to coordinate mobility, aggregation,
and inference so that environmental transitions inform network
organization and model selection while domain-specific knowledge
remains available across aggregation tiers. This motivates the
proposed framework under the modeled energy and latency constraints.
\subsection{Contributions}

The main contributions of this paper are as follows:

\begin{itemize}

\item We propose a mobility-based domain-drift measure that
estimates whether a \ac{UAV} will leave its current environmental
zone before the next synchronization event and combines this
estimate with entropy-based feedback and environment signatures
to guide clustering and \ac{CH} selection under energy and
latency constraints.

\item We propose a domain-aware architecture comprising a shared
backbone, environment-specific prediction heads, and local
normalization components, with local training integrating
domain-adversarial learning, proximal regularization, and
SCAFFOLD correction while keeping all normalization components
local to each \ac{UAV}.

\item We develop a two-tier aggregation method that applies
sample-weighted aggregation within clusters and Fisher-weighted
merging at the cloud, while additive head statistics preserve
the direct Fisher merge over accepted contributions and version
checks prevent incompatible model updates from being combined.

\item We develop a signature-based gate that selects an
environment-specific head or mixes the two most likely heads
when the selection is uncertain, while entropy-based drift
detection triggers limited adaptation of local normalization
without modifying the shared backbone.

\item We derive resource-feasibility conditions, a conditional
backbone-stationarity bound, and quadratic optimality of
Fisher-weighted merging, and analyze complexity and evaluate
classification, drift, clustering, and modeled energy use
through feature-level simulations and component ablations.

\end{itemize}

\section{System Model}
\begin{figure*}
\centering 
\resizebox{1.4\columnwidth}{!}{
\includegraphics{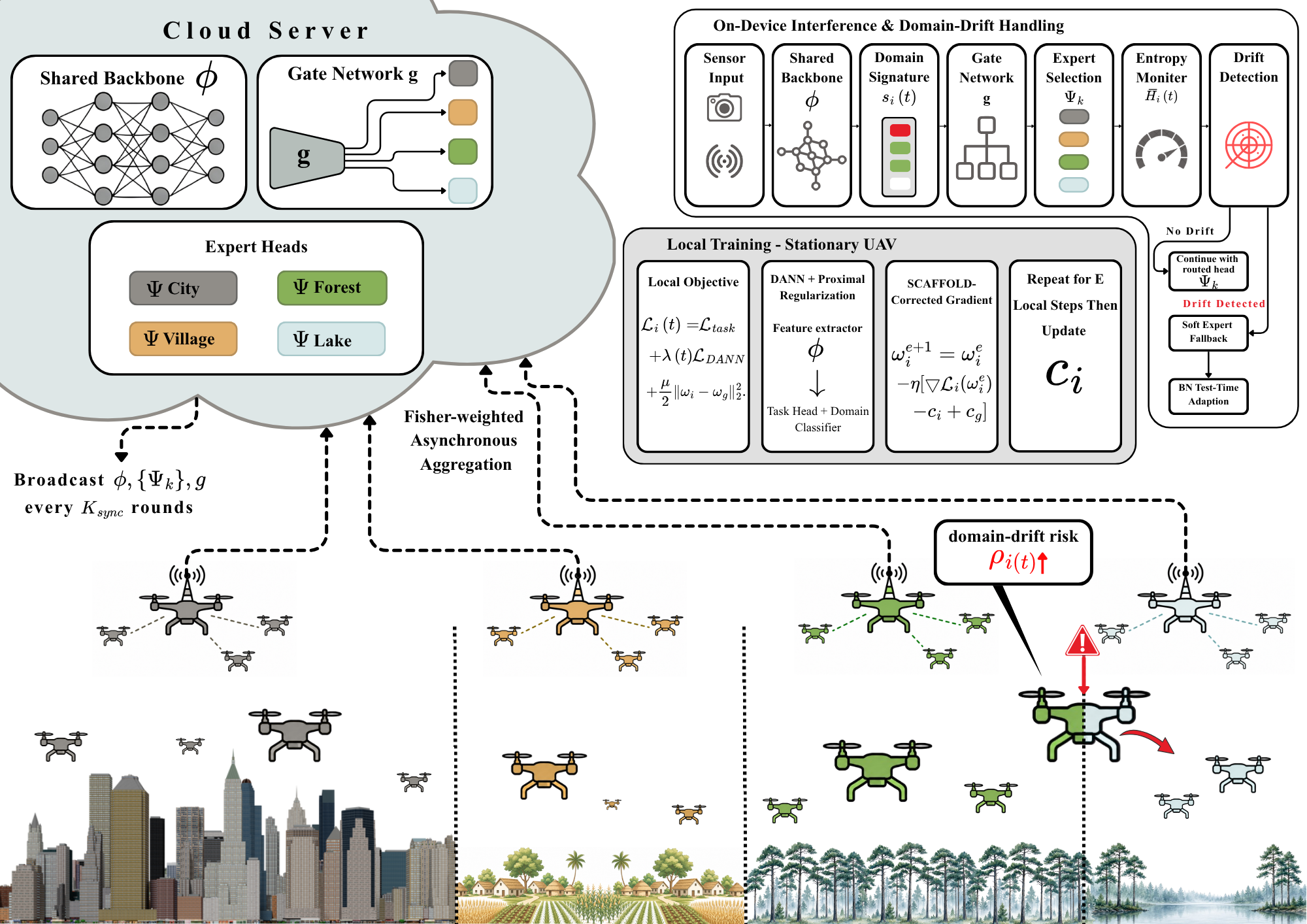}}
\caption{Overall architecture of the proposed domain-aware hierarchical federated learning framework for mobile UAV swarms.}
\label{fig:overall_architecture}
\end{figure*}

We consider an \ac{HFL} system comprising a \ac{UAV} swarm and cloud server, as shown in Fig.~\ref{fig:overall_architecture}. Each \ac{UAV} collects data and trains locally, while \acp{CH} aggregate and forward updates to the cloud. The swarm operates across zones with different data distributions, with \ac{UAV} mobility inducing distribution shifts.

%%%%%%%%%%%%%%%%%%%%%%%%%%%%%%%%%%%%%%%%%%%%%%%%%%%%%%%%%%%%%%%
%%%%%%%%%%%%%%%%%%%%%%%%%%%%%%%%%%%%%%%%%%%%%%%%%%%%%%%%%%%%%%

\subsection{Network Topology}
\label{sec:network_topology}

The network consists of a cloud server and a swarm
of \acp{UAV}. Let $\mathcal{N}=\{1,2,\ldots,N\}$
denote the UAV-index set, where $N$ is the total
number of \acp{UAV} and $i\in\mathcal{N}$ identifies
the $i$-th \ac{UAV}. The \ac{FL} process consists
of $T$ learning rounds indexed by
$t\in\mathcal{T}=\{0,1,\ldots,T-1\}$, where
$\mathcal{T}$ is the round-index set. Each round
involves local model training, model exchange,
and aggregation. At round $t$, the swarm is organized into $M_t$
nonempty clusters, where $M_t$ denotes the number
of active clusters. Let
$\mathcal{M}_t=\{1,2,\ldots,M_t\}$ denote their
index set and $\mathcal{C}_m^t\subseteq\mathcal{N}$
the set of UAV indices assigned to cluster
$m\in\mathcal{M}_t$. Since the clusters collectively
cover the entire swarm, their union satisfies $\bigcup_{m\in\mathcal{M}_t}\mathcal{C}_m^t=\mathcal{N}.$ Moreover, each \ac{UAV} belongs to only one cluster,
which requires the membership sets to satisfy
\begin{equation}
\mathcal{C}_m^t\cap\mathcal{C}_{m'}^t
=
\emptyset,
\quad
\forall m,m'\in\mathcal{M}_t,\;m\neq m',
\label{eq:cluster_partition}
\end{equation}
where $m'$ identifies a cluster different from $m$
and $\emptyset$ denotes the empty set. To represent these memberships in the subsequent problem formulation, the binary association variable
$x_{i,m}^t\in\{0,1\}$ is defined as
\begin{equation}
x_{i,m}^t=
\begin{cases}
1, & \text{if } i\in\mathcal{C}_m^t,\\
0, & \text{otherwise}.
\end{cases}
\label{eq:uav_cluster_association}
\end{equation}
Accordingly, the requirement that each \ac{UAV}
belongs to exactly one cluster is expressed as $\sum_{m\in\mathcal{M}_t}x_{i,m}^t=1, \forall i\in\mathcal{N}.$
The association variables form the binary assignment
matrix $\mathbf{X}^t=[x_{i,m}^t]
\in\{0,1\}^{N\times M_t}$, whose rows correspond
to \acp{UAV} and columns correspond to clusters.
The number of members in cluster $m$ at round $t$,
denoted by $n_m^t$, is obtained as
\begin{equation}
n_m^t
=
|\mathcal{C}_m^t|
=
\sum_{i\in\mathcal{N}}x_{i,m}^t,
\quad
\forall m\in\mathcal{M}_t,
\label{eq:cluster_size}
\end{equation}
where $|\mathcal{C}_m^t|$ denotes the number of
elements in $\mathcal{C}_m^t$. Each active cluster
is nonempty and therefore satisfies $n_m^t\geq1$. Each cluster selects one member as its \ac{CH}.
Let $h_m^t\in\mathcal{C}_m^t$ denote the index of
the selected \ac{CH} of cluster $m$ at round $t$.
The corresponding binary selection variable
$z_{i,m}^t$ is defined as
\begin{equation}
z_{i,m}^t=
\begin{cases}
1, & \text{if } i=h_m^t,\\
0, & \text{otherwise}.
\end{cases}
\label{eq:ch_indicator}
\end{equation}
Selecting exactly one \ac{CH} per cluster requires $\sum_{i\in\mathcal{N}}z_{i,m}^t=1,\;\forall m\in\mathcal{M}_t$, with cluster membership enforced by $z_{i,m}^t\leq x_{i,m}^t,\;\forall i\in\mathcal{N},\;m\in\mathcal{M}_t$. Each \ac{UAV} performs local training, while the \ac{CH} aggregates member updates and forwards the cluster contribution to the cloud for inter-cluster aggregation. Since \ac{CH} uploads are asynchronous, contributions may arrive at different times, with $t$ denoting their originating round. Cluster membership and \ac{CH} selection are updated during reclustering to reflect mobility and resource changes.

\subsection{Mobility and Data Model}
\label{sec:mobility_model}
\label{sec:data_model}

The position of each \ac{UAV} determines its occupied environmental zone. Let $\mathbf{p}_i(t),\mathbf{v}_i(t)\in\mathbb{R}^{3}$ denote the 3D position and velocity of the $i$-th \ac{UAV} at round $t$, respectively, and let $\Delta t>0$ denote the interval between successive mobility observations. With commanded acceleration $\mathbf{a}_i(t)\in\mathbb{R}^{3}$ assumed constant over this interval, the position evolves as
\begin{equation}
\mathbf{p}_i(t+1)
=\mathbf{p}_i(t)+\mathbf{v}_i(t)\Delta t
+\frac{1}{2}\mathbf{a}_i(t)(\Delta t)^2
+\mathbf{d}_i^{\mathrm{pos}}(t),
\label{eq:kinematics_updated}
\end{equation}
where $\mathbf{d}_i^{\mathrm{pos}}(t)\in\mathbb{R}^{3}$ represents the accumulated position disturbance due to wind and control errors. These disturbances are assumed independent across rounds and distributed as $\mathbf{d}_i^{\mathrm{pos}}(t)
\sim\mathcal{N}(\mathbf{0},\mathbf{C}_{\mathrm{pos}}),
\mathbf{C}_{\mathrm{pos}}
=\operatorname{diag}(\sigma_x^2,\sigma_y^2,\sigma_z^2),$ where $\mathcal{N}(\mathbf{0},\mathbf{C}_{\mathrm{pos}})$ denotes zero-mean Gaussian process noise with covariance matrix $\mathbf{C}_{\mathrm{pos}}\in\mathbb{R}^{3\times3}$, and $\sigma_x^2$, $\sigma_y^2$, and $\sigma_z^2$ are the per-interval position-error variances along the three coordinate axes. The trajectories are mission-specified and not optimized.

Let $\mathcal{A}\subseteq\mathbb{R}^{2}$ denote the ground-plane mission area, partitioned into $K$ known environmental zones indexed by $\mathcal{K}=\{1,\ldots,K\}$, with $\mathcal{Z}_k\subseteq\mathcal{A}$ denoting the $k$-th zone. The complete coverage and exclusive zone membership require
\begin{equation}
\mathcal{A}
=\bigcup_{k\in\mathcal{K}}\mathcal{Z}_k,
\mathcal{Z}_k\cap\mathcal{Z}_{k'}=\emptyset,
\quad k,k'\in\mathcal{K},\;k\neq k',
\label{eq:environment_partition}
\end{equation}
where $k'$ denotes a zone different from $k$, and boundary points are uniquely assigned. The ground-plane projection matrix $\Pi=[\,\mathbf{I}_2\ \mathbf{0}_{2\times1}\,]\in\mathbb{R}^{2\times3}$ retains horizontal coordinates and removes altitude. For positions projected within $\mathcal{A}$, the occupied-zone index is defined as $k(i,t)=k$
if and only if
$\Pi\mathbf{p}_i(t)\in\mathcal{Z}_k, k\in\mathcal{K}.$ A transition $k(i,t+1)\neq k(i,t)$ may change the local data distribution. For each zone $k\in\mathcal{K}$, let $\mathcal{D}_k$ denote the joint input--label distribution over $\mathcal{X}\times\mathcal{Y}$, where $\mathcal{X}$ is the input space and $\mathcal{Y}$ the set of class labels. Each zone distribution is assumed fixed during the mission. The local dataset collected by the $i$-th \ac{UAV} at round $t$ is defined as
\begin{equation}
\mathcal{S}_i(t)
=
\left\{
\bigl(\mathbf{x}_i^{(j)}(t),y_i^{(j)}(t)\bigr)
\right\}_{j=1}^{n_i^{\mathrm{data}}(t)},
\label{eq:local_dataset}
\end{equation}
where $n_i^{\mathrm{data}}(t)=|\mathcal{S}_i(t)|$ is the local sample count, $j\in\{1,\ldots,n_i^{\mathrm{data}}(t)\}$ is the sample index, and $\mathbf{x}_i^{(j)}(t)\in\mathcal{X}$ and $y_i^{(j)}(t)\in\mathcal{Y}$ denote the $j$-th input and its corresponding class label. Conditional on the occupied zone, samples within each round's dataset are assumed independent and identically distributed (i.i.d.) according to $\bigl(\mathbf{x}_i^{(j)}(t),y_i^{(j)}(t)\bigr)
\overset{\mathrm{i.i.d.}}{\sim}
\mathcal{D}_{k(i,t)}, j=1,\ldots,n_i^{\mathrm{data}}(t).$ This assumption does not require independence across rounds. \Acp{UAV} in the same zone share a population distribution but may collect different datasets. Different zone distributions yield spatial heterogeneity, while transitions between zones with $\mathcal{D}_k\neq\mathcal{D}_{k'}$ induce temporal shifts in input appearance, class frequencies, or both. To quantify these differences, let $P$ and $Q$ denote joint input--label distributions with densities $p(\mathbf{x},y)$ and $q(\mathbf{x},y)$, respectively, where $\mathbf{x}\in\mathcal{X}$ and $y\in\mathcal{Y}$. For continuous inputs and discrete labels, their \ac{K-L} divergence is defined as
\begin{equation}
\operatorname{KL}(P\Vert Q)
=
\sum_{y\in\mathcal{Y}}
\int_{\mathcal{X}}
p(\mathbf{x},y)
\log\frac{p(\mathbf{x},y)}{q(\mathbf{x},y)}
\,d\mathbf{x},
\label{eq:kl_definition}
\end{equation}
which is infinite if $P$ assigns positive probability to a set assigned zero probability by $Q$. For zones $k$ and $k'$, let $\bar{\mathcal{D}}_{k,k'}=\tfrac{1}{2}(\mathcal{D}_k+\mathcal{D}_{k'})$ denote their equally weighted mixture distribution. The corresponding \ac{JS} divergence is defined as
\begin{equation}
\operatorname{JS}(\mathcal{D}_k\Vert\mathcal{D}_{k'})
=
\frac{1}{2}\operatorname{KL}
\bigl(\mathcal{D}_k\Vert\bar{\mathcal{D}}_{k,k'}\bigr)
+
\frac{1}{2}\operatorname{KL}
\bigl(\mathcal{D}_{k'}\Vert\bar{\mathcal{D}}_{k,k'}\bigr).
\label{eq:js_divergence}
\end{equation}
This measure is symmetric, lies in $[0,\log 2]$, and equals zero if and only if the distributions coincide, characterizing population-level environmental dissimilarity. Since zone distributions are unknown, the framework identifies environments using signatures computed from local samples.

\subsection{Channel and Energy Model}
\label{sec:channel_energy}

The communication cost of each \ac{UAV} depends
on its role in the hierarchy. Cluster members
transmit local updates to their \ac{CH}, while
the \ac{CH} forwards the aggregated contribution
to the cloud. For a non-head member
$i\in\mathcal{C}_m^t\setminus\{h_m^t\}$, the
distance to its serving \ac{CH} is expressed as
$d_{i,m}(t)=
\left\|
\mathbf{p}_i(t)-\mathbf{p}_{h_m^t}(t)
\right\|_2,$
where $\mathbf{p}_{h_m^t}(t)$ is the position
of the selected \ac{CH}. The set difference
excludes the \ac{CH}, which incorporates its
own update directly into aggregation. The member-to-\ac{CH} \ac{A2A} channel coefficient is modeled
using distance-dependent large-scale attenuation and small-scale
Rician fading, following the standard UAV channel structure in
\cite{wang2020task}, as
\begin{equation}
h_{i,m}(t)
=
\sqrt{
\beta_0
\left(\frac{d_{i,m}(t)}{d_{\mathrm{ref}}}\right)^{-\alpha}
}
\,\widetilde{h}_{i,m}(t),
\label{eq:channel_model}
\end{equation}
where $d_{\mathrm{ref}}=1\,\mathrm{m}$ is the
reference distance, $\beta_0$ is the average
channel power gain at that distance, and
$\alpha$ is the path-loss exponent. The
coefficient $\widetilde{h}_{i,m}(t)$ represents
normalized Rician fading and satisfies
$\mathbb{E}[|\widetilde{h}_{i,m}(t)|^2]=1$. Its Rician factor $K_{\mathrm{ric}}$
is the ratio of specular to scattered power. Assuming negligible interference, the received
\ac{SNR} is
$
\mathrm{SNR}_{i,m}(t)
=
\frac{P_i^{\mathrm{tx}}|h_{i,m}(t)|^2}{N_oB},
$
where $P_i^{\mathrm{tx}}$ is the transmit power
of the $i$-th \ac{UAV}, $N_o$ is the receiver
noise \ac{PSD}, and $B$ is the allocated link
\ac{BW}. Accordingly, the Shannon transmission rate
is expressed as
$R_{i,m}(t)
=
B\log_2\bigl(1+\mathrm{SNR}_{i,m}(t)\bigr).$

Let $b_i(t)$ denote the payload size, including the local
update and accompanying learning information. Assuming
constant rate and transmit power, the transmission time
and energy are $T_i^{\mathrm{comm}}(t)=b_i(t)/R_{i,m}(t)$
and $E_i^{\mathrm{comm}}(t)=P_i^{\mathrm{tx}}T_i^{\mathrm{comm}}(t)$,
respectively. For cloud uploads, let $b_m^{\mathrm{up}}(t)$,
$R_m^{\mathrm{up}}(t)>0$, and $P_m^{\mathrm{up}}$ denote the
aggregated payload size, effective rate, and \ac{CH}
transmit power, respectively. The upload time and energy
are $T_m^{\mathrm{up}}(t)=b_m^{\mathrm{up}}(t)/R_m^{\mathrm{up}}(t)$
and $E_m^{\mathrm{up}}(t)=P_m^{\mathrm{up}}T_m^{\mathrm{up}}(t)$.
The cloud-link rate is treated as an input without applying
the \ac{A2A} propagation model. Thus, the transmit energy
of the $i$-th \ac{UAV} in cluster $m$ is defined as
\begin{equation}
E_i^{\mathrm{tx}}(t)
=
\begin{cases}
E_i^{\mathrm{comm}}(t), & i\neq h_m^t,\\
E_m^{\mathrm{up}}(t), & i=h_m^t.
\end{cases}
\label{eq:role_transmission_energy}
\end{equation}
The energy model accounts only for uplink transmission energy and does not consider reception, downlink transmission, or radio-circuit energy.

In addition to communication, each \ac{UAV}
consumes energy for local training. Let $E$
denote the number of training steps per round
and $\mathcal{B}_{i,e}(t)$ the mini-batch used
by the $i$-th \ac{UAV} at step
$e\in\{1,\ldots,E\}$. The number of processed
sample instances is calculated as $q_i(t)=\sum_{e=1}^{E}|\mathcal{B}_{i,e}(t)|,$
where $|\mathcal{B}_{i,e}(t)|$ is the mini-batch
size. Repeated use of a sample is counted
separately, so $q_i(t)$ may differ from the
local dataset size $n_i^{\mathrm{data}}(t)$.
Let $c_i$ denote the effective number of
\ac{CPU} cycles per processed training sample
and $f_i(t)>0$ denote the processor frequency. The local computation
time is expressed as
$T_i^{\mathrm{comp}}(t)=\frac{c_iq_i(t)}{f_i(t)}.$
Using the effective switched-capacitance model,
the computation energy is obtained as
$E_i^{\mathrm{comp}}(t)
=
c_{\mathrm{eff},i}c_iq_i(t)f_i(t)^2,$
where $c_{\mathrm{eff},i}$ is the effective
switched-capacitance coefficient of the
processor. 
These
expressions account for local training;
aggregation and inference workloads are
not explicitly modeled.
The propulsion cost is approximated using
the rotary-wing level-flight model in
\cite{zeng2019energy,dai2022energyefficient}.
Let $u_i(t)=\|\Pi\mathbf{v}_i(t)\|_2$ denote
the horizontal speed of the $i$-th \ac{UAV},
where $\Pi$ is the ground-plane projection
matrix defined in the mobility model.
Approximating this speed as constant over
$\Delta t$, the flight energy is expressed as
\begin{equation}
\begin{aligned}
E_i^{\mathrm{fly}}(t)
={}&[
P_0(1+\frac{3u_i(t)^2}{U_{\mathrm{tip}}^2})+P_1(
\sqrt{1+\frac{u_i(t)^4}{4v_0^4}}
-\\
&\frac{u_i(t)^2}{2v_0^2}
)^{1/2}+\frac{1}{2}d_od_{\mathrm{air}}sA\,u_i(t)^3
]\Delta t,
\end{aligned}
\label{eq:flight_energy}
\end{equation}
where $P_0$ and $P_1$ denote blade-profile and induced hover power, respectively. $U_{\mathrm{tip}}$ and $v_0$ are the rotor-tip speed and mean induced velocity, while $d_o$, $d_{\mathrm{air}}$, $s$, and $A$ denote the fuselage drag ratio, air density, rotor solidity, and rotor-disc area. The approximation excludes propulsion costs from vertical motion and acceleration. Since trajectories are mission-specified, flight energy is not optimized. The total modeled energy attributed to the $i$-th \ac{UAV} at round $t$ is given by
\begin{equation}
E_i(t)
=
E_i^{\mathrm{comp}}(t)
+
E_i^{\mathrm{tx}}(t)
+
E_i^{\mathrm{fly}}(t).
\label{eq:total_energy}
\end{equation}
The communication costs are assigned to the round
generating the update. Since uploads may
finish asynchronously, $E_i(t)$ represents
round-attributed energy rather than energy
necessarily consumed within the same
interval $\Delta t$.

\section{Problem Formulation}
\label{sec:problem_formulation}
Following the system model, the objective is to coordinate
swarm learning under energy and latency constraints while
addressing heterogeneity and mobility-induced distribution
shifts. Environment signatures and transition risks guide
clustering, while a shared backbone with environment-specific
heads supports learning across environments.

\subsection{Domain-Drift Risk}
\label{sec:drift-risk}

To guide clustering under mobility-induced
distribution changes, domain-drift risk
measures probability that the $i$-th \ac{UAV}
lies outside its current zone at the next
global synchronization.
Let $K_{\mathrm{sync}}$ denote the synchronization
interval in rounds and $H_t$ the remaining
rounds, with $1\leq H_t\leq K_{\mathrm{sync}}$.
Propagating \eqref{eq:kinematics_updated}
without disturbances gives the predicted mean position $\bar{\mathbf{p}}_i(t+H_t)$ as
\begin{equation} 
\mathbf{p}_i(t)+\sum_{r=0}^{H_t-1}
\Bigl[
\bar{\mathbf{v}}_i(t+r)\Delta t
+\tfrac{1}{2}\bar{\mathbf{a}}_i(t+r)(\Delta t)^2
\Bigr],
\label{eq:nominal_future_position}
\end{equation}
where $r$ indexes forecast intervals, and
$\bar{\mathbf{v}}_i(t+r)$ and
$\bar{\mathbf{a}}_i(t+r)$ are the nominal
velocity and acceleration.
Let $\mathcal{I}_i(t)$ contain the current
position and nominal motion sequence,
treated as known over the forecast horizon.
Conditional on this information, independent
position disturbances with distribution
$\mathcal{N}(\mathbf{0},\mathbf{C}_{\mathrm{pos}})$
give the future-position distribution as
$
\mathbf{p}_i(t+H_t)\mid\mathcal{I}_i(t)
\sim
\mathcal{N}\left(
\bar{\mathbf{p}}_i(t+H_t),
H_t\mathbf{C}_{\mathrm{pos}}
\right).
$
The ground-plane mean and covariance are
$\boldsymbol{\mu}_i(t)
=\Pi\bar{\mathbf{p}}_i(t+H_t)\in\mathbb{R}^{2}$
and
$\mathbf{S}_i(t)
=H_t\Pi\mathbf{C}_{\mathrm{pos}}\Pi^{\top}
\in\mathbb{R}^{2\times2}$, respectively.
The drift probability is defined by
$r_i(t)=
\Pr[\Pi\mathbf{p}_i(t+H_t)\notin
\mathcal{Z}_{k(i,t)}\mid\mathcal{I}_i(t)]$.
For positive horizontal position-error
variances, it is expressed as
\begin{equation}
r_i(t)
=
1-\int_{\mathcal{Z}_{k(i,t)}}
f_{\mathrm{G},2}\left(
\mathbf{u};
\boldsymbol{\mu}_i(t),\mathbf{S}_i(t)
\right)\,d\mathbf{u},
\label{eq:drift_risk_integral}
\end{equation}
where $\mathbf{u}\in\mathbb{R}^{2}$ is a
ground-plane location and $f_{\mathrm{G},2}$ denotes
the Gaussian density with the specified
mean and covariance.
The probability in \eqref{eq:drift_risk_integral}
is estimated using $J_{\mathrm{mc}}$ independent
Monte Carlo samples
$\mathbf{u}_i^{(j)}(t)\sim
\mathcal{N}(\boldsymbol{\mu}_i(t),\mathbf{S}_i(t))$,
where $j=1,\ldots,J_{\mathrm{mc}}$ indexes
the sampled ground positions.
The estimated probability is expressed as
\begin{equation}
\hat{r}_i(t)
=
\frac{1}{J_{\mathrm{mc}}}
\sum_{j=1}^{J_{\mathrm{mc}}}
\mathbb{1}\left[
\mathbf{u}_i^{(j)}(t)\notin\mathcal{Z}_{k(i,t)}
\right],
\label{eq:drift_monte_carlo}
\end{equation}
where $\mathbb{1}[\cdot]$ equals one when
the enclosed condition holds and zero otherwise. To incorporate recent detector feedback, let
$d_i(t')\in\{0,1\}$ denote the entropy-based
drift flag at round $t'$, with one indicating
detected drift. The preceding observation
window is
$\mathcal{W}_t=\{t'\in\mathcal{T}:
\max(0,t-K_{\mathrm{sync}})\leq t'<t\}$.
The adjusted risk score is defined as
\begin{equation} \small
\begin{aligned}
\tilde{r}_i(t)
=\min\Bigl\{1,\,
r_i(t)+a_{\mathrm{risk}}
\mathbb{1}\bigl[
\exists t'\in\mathcal{W}_t:
d_i(t')=1
\bigr]\Bigr\},
\end{aligned}
\label{eq:flagged_risk}
\end{equation}
where $a_{\mathrm{risk}}\in(0,1]$ controls the
risk increase following detected drift.
In implementation, $\hat{r}_i(t)$ replaces
$r_i(t)$ in \eqref{eq:flagged_risk}.
The resulting score guides clustering
and \ac{CH} eligibility.

\subsection{Model Decomposition and Environment Signature}
\label{sec:signature}

To retain knowledge shared across environments while
preserving environment-specific information, the
learning model comprises a shared backbone,
environment-specific prediction heads, and local
batch-normalization components. The complete
collection of parameters and statistics is denoted
$
\theta
=
\left(
\mathbf{w}_{\mathrm{back}},\{\psi_k\}_{k\in\mathcal{K}},
\{\mathbf{b}_i^{\mathrm{BN}},\xi_i\}_{i\in\mathcal{N}}
\right),
$
where $\mathbf{w}_{\mathrm{back}}\in\mathbb{R}^{d_{\mathrm{back}}}$ contains
shared backbone parameters and
$\psi_k\in\mathbb{R}^{d_\psi}$ contains the prediction
head parameters for zone $k$. The quantities
$d_{\mathrm{back}}$ and $d_\psi$ denote the numbers of backbone
and head parameters, respectively. For $i$-th
\ac{UAV}, $\mathbf{b}_i^{\mathrm{BN}}$ contains local normalization
statistics and $\xi_i$ contains trainable
normalization parameters. These local components
are excluded from $\mathbf{w}_{\mathrm{back}}$ and $\psi_k$.

Let $L_{\mathrm{bn}}$ denote the number of
batch-normalization layers and
$\ell\in\{1,\ldots,L_{\mathrm{bn}}\}$ their index.
The local statistics and trainable parameters
are defined as
\begin{equation}
\begin{aligned}
\mathbf{b}_i^{\mathrm{BN}}
&=
\left\{
\boldsymbol{\mu}_i^{(\ell)},
\mathbf{v}_i^{\mathrm{BN},(\ell)}
\right\}_{\ell=1}^{L_{\mathrm{bn}}},&
\xi_i
=
\left\{
\boldsymbol{\gamma}_i^{(\ell)},
\boldsymbol{\eta}_i^{(\ell)}
\right\}_{\ell=1}^{L_{\mathrm{bn}}},
\end{aligned}
\label{eq:bn_statistics}
\end{equation}
where $\boldsymbol{\mu}_i^{(\ell)}$ and $\mathbf{v}_i^{\mathrm{BN},(\ell)}$ denote the running mean and variance vectors, while $\boldsymbol{\gamma}_i^{(\ell)}$ and $\boldsymbol{\eta}_i^{(\ell)}$ are the trainable scale and offset parameters at layer $\ell$, with one entry per normalized feature channel. The backbone is trained across environments, whereas each head uses data from its corresponding zone. Following the local-normalization principle of \ac{FedBN}, $\mathbf{b}_i^{\mathrm{BN}}$ and $\xi_i$ remain local and are not aggregated.

Since the population distributions $\mathcal{D}_k$ are unknown during operation, the \ac{JS} divergence in \eqref{eq:js_divergence} cannot be evaluated directly. Instead, the framework uses an environment signature computed from local input samples. Let $\mathcal{B}_i(t)\subseteq\mathcal{S}_i(t)$ denote a nonempty mini-batch collected by the $i$-th \ac{UAV}, with input dimension $d_x$ and $\mathbf{x}\in\mathbb{R}^{d_x}$. The empirical input mean is
$
\boldsymbol{\mu}_i^x(t)
=
\frac{\sum_{(\mathbf{x},y)\in\mathcal{B}_i(t)}
\mathbf{x}}{|\mathcal{B}_i(t)|},
$
where $|\mathcal{B}_i(t)|$ is mini-batch size and $\boldsymbol{\mu}_i^x(t)\in\mathbb{R}^{d_x}$ contains mean of each input component. The corresponding componentwise empirical variance is
\begin{equation}
\mathbf{v}_i^{x}(t)
=
\frac{1}{|\mathcal{B}_i(t)|}
\sum_{(\mathbf{x},y)\in\mathcal{B}_i(t)}
\left(
\mathbf{x}-\boldsymbol{\mu}_i^x(t)
\right)^{\odot2},
\label{eq:signature_variance}
\end{equation}
where $\mathbf{v}_i^{x}(t)
\in\mathbb{R}^{d_x}$ contains the componentwise
variances and $(\cdot)^{\odot2}$ is
elementwise squaring. Although the mini-batch
contains input--label pairs, these statistics
use only the inputs.
The environment signature is formed by
concatenating the mean and variance vectors as
$
\mathbf{s}_i(t)
=
\begin{bmatrix}
\boldsymbol{\mu}_i^x(t)\\
\mathbf{v}_i^{x}(t)
\end{bmatrix}
\in\mathbb{R}^{2d_x},
$
where $\mathbf{s}_i(t)$ summarizes the observed
input distribution of the $i$-th \ac{UAV}.
Since the signature is computed from raw inputs,
its calculation does not depend on the backbone
parameters. The signature supports environment
similarity clustering and prediction-head selection,
while the adjusted risk $\tilde{r}_i(t)$
accounts for anticipated transitions and recent
drift observations.

\subsection{Per-Environment and Aggregate Objectives}
\label{sec:learning_objectives}

Following model decomposition, the learning objective trains the
shared backbone and environment-specific prediction heads while
accounting for local normalization. Let
$\ell(\mathbf{x},y;\mathbf{w}_{\mathrm{back}},\psi_k,\mathbf{b}_i^{\mathrm{BN}},\xi_i)\geq0$
be supervised loss for input $\mathbf{x}$ and label $y$,
evaluated using $\mathbf{w}_{\mathrm{back}}$, $\psi_k$, $\mathbf{b}_i^{\mathrm{BN}}$ and $\xi_i$ of the $i$-th
\ac{UAV}. For zone $k$, the population risk under the $i$-th
\ac{UAV}'s normalization components is defined as $
F_k^{(i)}(\mathbf{w}_{\mathrm{back}},\psi_k;\mathbf{b}_i^{\mathrm{BN}},\xi_i)
=
\mathbb{E}_{(\mathbf{x},y)\sim\mathcal{D}_k}
\left[
\ell(\mathbf{x},y;\mathbf{w}_{\mathrm{back}},\psi_k,\mathbf{b}_i^{\mathrm{BN}},\xi_i)
\right],
$ where $\mathcal{D}_k$ denotes joint input--label distribution
of zone $k$. Since $\mathcal{D}_k$ is unknown, the $i$-th \ac{UAV}
estimates this risk from its local samples. For its occupied zone
$k=k(i,t)$, the empirical risk $\widehat{F}_k^{(i,t)} 
(\mathbf{w}_{\mathrm{back}},\psi_k;\mathbf{b}_i^{\mathrm{BN}},\xi_i)$ is
\begin{equation}
\frac{1}{n_i^{\mathrm{data}}(t)}
\sum_{j=1}^{n_i^{\mathrm{data}}(t)}
\ell\bigl(
\mathbf{x}_i^{(j)}(t),y_i^{(j)}(t);
\mathbf{w}_{\mathrm{back}},\psi_k,\mathbf{b}_i^{\mathrm{BN}},\xi_i
\bigr),
\label{eq:local_empirical_risk}
\end{equation}
where $n_i^{\mathrm{data}}(t)>0$ is the local sample count. This supervised
objective is optimized using mini-batch gradients together with the
regularization and correction terms introduced later.

To characterize swarm-wide performance, let
$\mathbf{b}^{\mathrm{BN}}
=(\mathbf{b}_i^{\mathrm{BN}})_{i\in\mathcal{N}}$
and
$\boldsymbol{\xi}
=(\xi_i)_{i\in\mathcal{N}}$ denote the collections of
local normalization statistics and trainable parameters. The
population risk for zone $k$ is defined as $
F_k(\mathbf{w}_{\mathrm{back}},\psi_k;\mathbf{b}^{\mathrm{BN}},\boldsymbol{\xi})
=
\frac{1}{N}
\sum_{i\in\mathcal{N}}
F_k^{(i)}(\mathbf{w}_{\mathrm{back}},\psi_k;\mathbf{b}_i^{\mathrm{BN}},\xi_i).
$ Let $\pi_k\geq0$ denote the mission weight assigned to zone $k$,
with $\sum_{k\in\mathcal{K}}\pi_k=1$, and let
$\boldsymbol{\psi}=(\psi_1,\ldots,\psi_K)$ denote the collection of
prediction heads, the aggregate learning objective is defined as
\begin{equation} \small
\mathcal{J}
(\mathbf{w}_{\mathrm{back}},\boldsymbol{\psi};
\mathbf{b}^{\mathrm{BN}},\boldsymbol{\xi})=
\sum_{k\in\mathcal{K}}
\pi_k
F_k(\mathbf{w}_{\mathrm{back}},\psi_k;\mathbf{b}^{\mathrm{BN}},\boldsymbol{\xi}),
\label{eq:joint_learning_objective}
\end{equation}
where $\mathcal{J}$ is the mission-weighted population risk. For a fixed backbone and normalization components, an optimal head
for zone $k$ is characterized by $
\psi_k^\star
(\mathbf{w}_{\mathrm{back}};\mathbf{b}^{\mathrm{BN}},\boldsymbol{\xi})
\in
\operatorname*{arg\,min}_{\psi_k}
F_k(\mathbf{w}_{\mathrm{back}},\psi_k;\mathbf{b}^{\mathrm{BN}},\boldsymbol{\xi}).
$ Substituting these heads gives the reduced backbone objective, $F(\mathbf{w}_{\mathrm{back}};
\mathbf{b}^{\mathrm{BN}},\boldsymbol{\xi})$ as
\begin{equation}
\sum_{k\in\mathcal{K}}\pi_k F_k\Bigl(
\mathbf{w}_{\mathrm{back}},
\psi_k^\star(\mathbf{w}_{\mathrm{back}};
\mathbf{b}^{\mathrm{BN}},\boldsymbol{\xi});
\mathbf{b}^{\mathrm{BN}},\boldsymbol{\xi}
\Bigr).
\label{eq:aggregate_objective}
\end{equation}
This objective gives the population risk with optimal heads
for a fixed backbone and local normalization state.
The method updates the model iteratively without exact
head minimization.

\subsection{Cluster-Level Federated Aggregation}
\label{sec:cluster_aggregation}

Following local training, each \ac{CH} combines the
backbone updates produced by its cluster members.
Let $\mathbf{w}_{\mathrm{back},i}^{t}\in\mathbb{R}^{d_{\mathrm{back}}}$ denote the
backbone parameters of $i$-th \ac{UAV} after
local training at round $t$. To account for
differences in local dataset sizes, the contribution
of each member is weighted by its sample count.
For $i\in\mathcal{C}_m^t$, aggregation weight
is defined as
\begin{equation}
w_{i,m}^t
=
{n_i^{\mathrm{data}}(t)}/
{\displaystyle\sum_{j\in\mathcal{C}_m^t}n_j^{\mathrm{data}}(t)},
\label{eq:aggregation_weights}
\end{equation}
where $w_{i,m}^t$ is the weight assigned to the
$i$-th \ac{UAV}, $n_i^{\mathrm{data}}(t)$ is its local sample
count, and $j$ indexes the \acp{UAV} in cluster
$m$. Assuming positive sample counts, these
weights are nonnegative and satisfy
$\sum_{i\in\mathcal{C}_m^t}w_{i,m}^t=1$.
Using these weights, the aggregated backbone
of cluster $m$ is obtained as
\begin{equation}
\mathbf{w}_{\mathrm{back},m}^{t}
=
\sum_{i\in\mathcal{C}_m^t}w_{i,m}^t\mathbf{w}_{\mathrm{back},i}^{t},
\label{eq:cluster_aggregation}
\end{equation}
where $\mathbf{w}_{\mathrm{back},m}^{t}\in\mathbb{R}^{d_{\mathrm{back}}}$ denotes the cluster-level backbone parameters, including the \ac{CH}'s local update since $h_m^t\in\mathcal{C}_m^t$. The weighting follows the sample-based aggregation principle of \cite{mcmahan2017communication}. The corresponding cluster-level empirical objective is
$
\mathcal{F}_m^t
(\mathbf{w}_{\mathrm{back}},\boldsymbol{\psi};
\mathbf{b}^{\mathrm{BN}},\boldsymbol{\xi})
=
\sum_{i\in\mathcal{C}_m^t}
w_{i,m}^t
\widehat{F}_{k(i,t)}^{(i,t)}
\bigl(\mathbf{w}_{\mathrm{back}},\psi_{k(i,t)};\mathbf{b}_i^{\mathrm{BN}},\xi_i\bigr),
$
where $\mathcal{F}_m^t$ is the weighted empirical risk, $\boldsymbol{\psi}$ contains environment-specific heads, and $\mathbf{b}^{\mathrm{BN}}$ and $\boldsymbol{\xi}$ collect the local normalization statistics and trainable parameters, respectively. Each member's loss uses the shared backbone, its zone-specific
head, and local normalization.
Equation~\eqref{eq:cluster_aggregation} averages trained backbones
without exactly minimizing the cluster objective.
Heads are grouped by zone as
$
\mathcal{C}_{m,k}^t
=\{i\in\mathcal{C}_m^t:k(i,t)=k\}.
$
Each nonempty group follows the Fisher-weighted rule in
\eqref{eq:fisher_merge}. Heads remain separate across zones,
while $\mathbf{b}_i^{\mathrm{BN}}$ and $\xi_i$ remain local
and are not transmitted. The \ac{CH} forwards the aggregated
backbone and zone-specific head contributions to the cloud.

\subsection{Learning Objective and Resource Constraints}
\label{sec:main_problem}
Based on the learning objectives and aggregation
structure, the framework trains the shared backbone
and environment-specific heads while maintaining
a resource-feasible network configuration.
For fixed local normalization components
$\mathbf{b}^{\mathrm{BN}}$ and $\boldsymbol{\xi}$,
the reference population learning objective is
\begin{equation}
\min_{\mathbf{w}_{\mathrm{back}},\boldsymbol{\psi}}
\mathcal{J}
(\mathbf{w}_{\mathrm{back}},\boldsymbol{\psi};
\mathbf{b}^{\mathrm{BN}},\boldsymbol{\xi}),
\label{eq:main_objective}
\end{equation}
where $\mathcal{J}$ is the mission-weighted population risk in \eqref{eq:joint_learning_objective}. At round $t$, environmental similarity and drift risk guide clustering and \ac{CH} selection, while energy and latency constraints govern \ac{CPU} frequency allocation. Let $E_i^{\max}$ and $f_i^{\max}>0$ denote the per-round energy budget and maximum \ac{CPU} frequency of the $i$-th \ac{UAV}. The energy cost $E_i(t)$ in \eqref{eq:total_energy} includes training, role-dependent transmission, propulsion, and asynchronous uploads attributed to the round. Assuming successive training, member-transmission, and cloud-upload stages with parallel member transmissions, the cluster completion time is
$
L_m(t)
=
\max_{i\in\mathcal{C}_m^t}
T_i^{\mathrm{comp}}(t)
+
\max_{i\in\mathcal{C}_m^t\setminus\{h_m^t\}}
T_i^{\mathrm{comm}}(t)
+
T_m^{\mathrm{up}}(t).
$
The member-transmission term is zero for a cluster containing only its \ac{CH}; aggregation processing and additional queueing delays are neglected. Let $L_{\max}>0$ denote the cluster completion deadline from the round's start, allowing asynchronous arrivals, and $r_{\mathrm{thresh}}\in[0,1]$ the maximum adjusted drift risk permitted for a \ac{CH}. The scores $\tilde{r}_i(t)$ in \eqref{eq:flagged_risk} are evaluated before assignment and held fixed during the network decision. For practical network organization, environmental similarity
and leadership drift risk are combined into a weighted scalar
utility. Let $\lambda_{\mathrm{sim}},\lambda_{\mathrm{risk}}\geq0$
satisfy
$\lambda_{\mathrm{sim}}+\lambda_{\mathrm{risk}}=1$.
At each reclustering event, the network configuration is obtained
from
\begin{equation} \small
\begin{aligned}
\max_{\mathbf{X}^t,\mathbf{Z}^t,\mathbf{f}(t)}
\quad
&\mathcal{Q}_t(\mathbf{X}^t,\mathbf{Z}^t)
=
\frac{\lambda_{\mathrm{sim}}}{N}
\sum_{i\in\mathcal{N}}
\sum_{m\in\mathcal{M}_t}
x_{i,m}^t\Gamma_{i,m}^{\mathrm{env}}(t)
\\
&-
\frac{\lambda_{\mathrm{risk}}}{M_t}
\sum_{m\in\mathcal{M}_t}
\sum_{i\in\mathcal{N}}
z_{i,m}^t\widetilde{r}_i(t),
\end{aligned}
\label{eq:cluster_score}
\end{equation}
\begin{subequations} \small
\label{eq:main_problem}
\begin{align}
\mathrm{s.t.}\quad
&E_i(t)\leq E_i^{\max},
&&\forall i\in\mathcal{N},
\label{eq:energy_constraint}\\
&0<f_i(t)\leq f_i^{\max},
&&\forall i\in\mathcal{N},
\label{eq:frequency_constraint}\\
&\sum_{m\in\mathcal{M}_t}x_{i,m}^t=1,
&&\forall i\in\mathcal{N},
\label{eq:assignment_constraint}\\
&\sum_{i\in\mathcal{N}}x_{i,m}^t\geq1,
&&\forall m\in\mathcal{M}_t,
\label{eq:nonempty_constraint}\\
&\sum_{i\in\mathcal{N}}z_{i,m}^t=1,
&&\forall m\in\mathcal{M}_t,
\label{eq:single_head_constraint}\\
&z_{i,m}^t\leq x_{i,m}^t,
&&\forall i,m,
\label{eq:head_membership_constraint}\\
&x_{i,m}^t,z_{i,m}^t\in\{0,1\},
&&\forall i,m,
\label{eq:binary_constraint}\\
&L_m(t)\leq L_{\max},
&&\forall m\in\mathcal{M}_t,
\label{eq:latency_constraint}\\
&\sum_{i\in\mathcal{N}}
z_{i,m}^t\widetilde{r}_i(t)
\leq r_{\mathrm{thresh}},
&&\forall m\in\mathcal{M}_t.
\label{eq:risk_constraint}
\end{align}
\end{subequations}
Constraint \eqref{eq:energy_constraint} limits each \ac{UAV}'s energy consumption, while \eqref{eq:frequency_constraint} bounds its \ac{CPU} frequency. Constraints \eqref{eq:assignment_constraint} and \eqref{eq:nonempty_constraint} enforce unique cluster assignment and nonempty active clusters. Constraints \eqref{eq:single_head_constraint} and \eqref{eq:head_membership_constraint} select one \ac{CH} per cluster, with \eqref{eq:binary_constraint} defining binary decisions. Constraint \eqref{eq:latency_constraint} limits cluster completion time, while \eqref{eq:risk_constraint} bounds the selected \ac{CH}'s drift-risk score by $r_{\mathrm{thresh}}\in[0,1]$. The objective in \eqref{eq:main_objective} defines the learning target, while \eqref{eq:main_problem} specifies feasible configurations.

\subsection{Lagrangian Formulation and Stationarity}
\label{sec:lagrangian}

For a fixed feasible cluster configuration,
the continuous variables are $\mathbf{w}_{\mathrm{back}}$,
$\boldsymbol{\psi}$, and $\mathbf{f}(t)$.
The normalization components, training workloads,
payloads, link rates, and propulsion costs
are held fixed.
Using the training workload $C_i(t)=c_iq_i(t)>0$,
let $e_i(t)=E_i^{\mathrm{tx}}(t)
+E_i^{\mathrm{fly}}(t)$ denote the
frequency-independent energy cost.
The energy constraint is expressed as
$
c_{\mathrm{eff},i}C_i(t)f_i(t)^2
+e_i(t)-E_i^{\max}\leq0.
$
The communication time of cluster $m$ is
defined as
\begin{equation}
a_m(t)=
\max_{i\in\mathcal{C}_m^t\setminus\{h_m^t\}}
T_i^{\mathrm{comm}}(t)
+T_m^{\mathrm{up}}(t),
\label{eq:fixed_communication_time}
\end{equation}
where the first term is zero for a singleton
cluster. Since $a_m(t)$ is fixed, the cluster
deadline is equivalently expressed through
the member-level constraints as
$
\frac{C_i(t)}{f_i(t)}+a_m(t)-L_{\max}\leq0,
\quad
\forall i\in\mathcal{C}_m^t,\;
m\in\mathcal{M}_t.
$
Let $\lambda_i$, $\lambda_i^{\mathrm{freq}}$, and $\eta_i$ denote
the nonnegative Lagrange multipliers for the
energy, upper-frequency, and latency constraints,
respectively. Writing $f_i=f_i(t)$,
$C_i=C_i(t)$, $e_i=e_i(t)$, and $a_m=a_m(t)$,
the Lagrangian is defined as
{\small
\begin{equation} 
\begin{split}
\mathcal{L}
={}\mathcal{J}(\mathbf{w}_{\mathrm{back}},\boldsymbol{\psi};
\mathbf{b}^{\mathrm{BN}},\boldsymbol{\xi})
+\sum_{i\in\mathcal{N}}\lambda_i
\left(
c_{\mathrm{eff},i}C_if_i^2
+e_i-E_i^{\max}
\right)\\
{}+\sum_{i\in\mathcal{N}}\lambda_i^{\mathrm{freq}}
\left(f_i-f_i^{\max}\right)
+\sum_{m\in\mathcal{M}_t}
\sum_{i\in\mathcal{C}_m^t}\eta_i
\left(
\frac{C_i}{f_i}+a_m-L_{\max}
\right).
\end{split}
\label{eq:lagrangian}
\end{equation}}
The domain requires $f_i>0$, while the fixed
configuration satisfies the membership,
\ac{CH}-selection, and risk constraints.
For a differentiable objective and under
a suitable constraint qualification, a local
minimizer satisfies the Karush--Kuhn--Tucker
conditions. A superscript $\dagger$ denotes
the candidate solution and its multipliers.
Backbone stationarity is expressed as
$
\sum_{k\in\mathcal{K}}\pi_k\nabla_{\mathbf{w}_{\mathrm{back}}}
F_k(\mathbf{w}_{\mathrm{back}}^\dagger,\psi_k^\dagger;
\mathbf{b}^{\mathrm{BN}},\boldsymbol{\xi})
=\mathbf{0},
$
where $\nabla_{\mathbf{w}_{\mathrm{back}}}$ denotes the backbone
gradient and $\mathbf{0}$ is the zero vector
of the corresponding dimension.
Stationarity for each head is expressed as
$
\pi_k\nabla_{\psi_k}
F_k(\mathbf{w}_{\mathrm{back}}^\dagger,\psi_k^\dagger;
\mathbf{b}^{\mathrm{BN}},\boldsymbol{\xi})
=\mathbf{0},
\quad \forall k\in\mathcal{K},
$
where $\nabla_{\psi_k}$ denotes the gradient
with respect to the $k$-th head.
For $\pi_k=0$, this condition is
identically satisfied.
Differentiating \eqref{eq:lagrangian}
with respect to $f_i$ gives
$
2\lambda_i^\dagger
c_{\mathrm{eff},i}C_if_i^\dagger
+\lambda_i^{\mathrm{freq},\dagger}
-\eta_i^\dagger\frac{C_i}{(f_i^\dagger)^2}
=0,
\quad \forall i\in\mathcal{N}.
$
The complementary-slackness conditions
are expressed as
\begin{align}
\lambda_i^\dagger
\left(
c_{\mathrm{eff},i}C_i(f_i^\dagger)^2
+e_i-E_i^{\max}
\right)
&=0,
\label{eq:comp_energy}\\
\lambda_i^{\mathrm{freq},\dagger}(f_i^\dagger-f_i^{\max})
&=0,
\label{eq:comp_upper_frequency}\\
\eta_i^\dagger
\left(
\frac{C_i}{f_i^\dagger}+a_m-L_{\max}
\right)
&=0,
\quad i\in\mathcal{C}_m^t,
\label{eq:comp_latency}
\end{align}
for all applicable indices, together with
primal feasibility and
$\lambda_i^\dagger,\lambda_i^{\mathrm{freq},\dagger},\eta_i^\dagger\geq0$.
These conditions do not guarantee global optimality for the nonconvex objective. Since $\mathcal{J}$ is independent of $\mathbf{f}(t)$, any feasible frequency allocation satisfies stationarity with zero resource multipliers; hence, energy-minimizing allocation requires a separate selection rule.
\subsection{Feasibility}
\label{sec:feasibility}

Following the fixed-configuration analysis, a
feasible resource allocation requires each
\ac{UAV} to complete local training within
the available time and energy budgets.
Consider a cluster configuration satisfying
the assignment, \ac{CH}-selection, and risk
constraints in \eqref{eq:main_problem}.
Using the communication time $a_m(t)$ defined
in \eqref{eq:fixed_communication_time}, the
time available for local computation in
cluster $m$ is expressed as
$
D_m(t)=L_{\max}-a_m(t),
$
where $D_m(t)$ is measured in seconds.
Since the training workload $C_i(t)$ is
positive, feasibility requires $D_m(t)>0$.
The minimum \ac{CPU} frequency required
to satisfy the latency constraint is
therefore obtained as
\begin{equation}
f_{i,\min}(t)
=
\frac{C_i(t)}{D_m(t)},
\qquad i\in\mathcal{C}_m^t,
\label{eq:minimum_feasible_frequency}
\end{equation}
where $f_{i,\min}(t)$ is the lowest frequency
at which the $i$-th \ac{UAV} can complete
its training within the available time.
The energy available for local training
is obtained by subtracting the transmission
and propulsion costs from assigned
budget as
$
E_i^{\mathrm{avail}}(t)
=
E_i^{\max}-e_i(t),
$
where $E_i^{\mathrm{avail}}(t)$ is
remaining training-energy allowance and
$e_i(t)$ is defined in preceding
subsection. For $c_{\mathrm{eff},i}>0$,
feasibility requires
$E_i^{\mathrm{avail}}(t)>0$.
The frequency upper bound imposed by the
energy constraint
\begin{equation}
f_{i,E}(t)
=
\sqrt{
\frac{E_i^{\mathrm{avail}}(t)}
{c_{\mathrm{eff},i}C_i(t)}
},
\label{eq:energy_frequency_limit}
\end{equation}
where $f_{i,E}(t)$ is the maximum frequency
permitted by the training-energy allowance.
Combining this bound with the hardware
limit, the maximum feasible frequency
is obtained as
$
f_{i,\max}^{\mathrm{feas}}(t)
=
\min\left\{
f_i^{\max},f_{i,E}(t)
\right\},
$
where $f_{i,\max}^{\mathrm{feas}}(t)$ is
the largest frequency satisfying both limits.
Accordingly, the fixed cluster configuration
admits a feasible frequency allocation if
and only if the following conditions hold
\begin{equation}
\begin{aligned}
&D_m(t)>0,\ \forall m\in\mathcal{M}_t,
\quad E_i^{\mathrm{avail}}(t)>0,\ \forall i\in\mathcal{N},\\
&f_{i,\min}(t)\leq f_{i,\max}^{\mathrm{feas}}(t),
\quad \forall i\in\mathcal{N}.
\end{aligned}
\label{eq:resource_feasibility}
\end{equation}
When these conditions hold, any frequency
within
$[f_{i,\min}(t),f_{i,\max}^{\mathrm{feas}}(t)]$
satisfies the modeled latency, energy,
and hardware constraints for the $i$-th
\ac{UAV}. In particular, selecting
$f_i(t)=f_{i,\min}(t)$ provides a feasible
allocation and minimizes local training
energy for the fixed workload and
cluster configuration.
These conditions ensure resource feasibility for a given assignment. Overall feasibility additionally requires at least one cluster configuration satisfying the membership and \ac{CH}-risk constraints along with \eqref{eq:resource_feasibility}.

\section{Proposed Methodology}
\label{sec:method}
The proposed \ac{DA-HFL} framework integrates environment-aware training, resource-feasible clustering, hierarchical aggregation, and local adaptation. The objective in \eqref{eq:main_objective} defines the learning target, while \eqref{eq:main_problem} specifies feasible configurations. Regularized updates, sample-weighted cluster backbones, and Fisher-weighted cloud aggregation enable learning without assuming exact minimization or stationarity.

\subsection{Local Model Training}
\label{sec:local_training}

At round $t$, the $i$-th \ac{UAV} trains the head associated with its occupied zone $k=k(i,t)$. Let $\mathbf{w}_{\mathrm{back},i}^{\mathrm{ref}}(t)$, $\psi_{i,k}^{\mathrm{ref}}(t)$, and $\mathbf{w}_{\mathrm{dom},i}^{\mathrm{ref}}(t)$ denote the latest received backbone, zone head, and domain discriminator parameters from one synchronized model version. The local backbone and head after $e$ steps are $\mathbf{w}_{\mathrm{back},i,e}^{t}$ and $\psi_{i,k,e}^t$, respectively, initialized as
$
\mathbf{w}_{\mathrm{back},i,0}^{t}=\mathbf{w}_{\mathrm{back},i}^{\mathrm{ref}}(t),
\qquad
\psi_{i,k,0}^t=\psi_{i,k}^{\mathrm{ref}}(t).
$
The normalization parameters retain their local initialization $\xi_{i,0}^t=\xi_i$. Consistent with the workload definition in Section~\ref{sec:channel_energy}, local training comprises $E$ steps indexed by $e=1,\ldots,E$, where step $e$ updates state $e-1$ using $\mathcal{B}_{i,e}(t)$. For local parameters $\mathbf{w}_{\mathrm{back},i}$ and $\psi_{i,k}$, the supervised mini-batch loss is
$
\mathcal{L}_{i,e}^{\mathrm{sup}}
(\mathbf{w}_{\mathrm{back},i},\psi_{i,k};\mathbf{b}_i^{\mathrm{BN}},\xi_i)
=
\frac{1}{|\mathcal{B}_{i,e}(t)|}
\sum_{(\mathbf{x},y)\in\mathcal{B}_{i,e}(t)}
\ell(\mathbf{x},y;\mathbf{w}_{\mathrm{back},i},\psi_{i,k},\mathbf{b}_i^{\mathrm{BN}},\xi_i).
$
To encourage shared features across zones through \ac{DANN}, a domain discriminator $D(\cdot;\mathbf{w}_{\mathrm{dom}})$ assigns probabilities to the $K$ environments. Let $g(\mathbf{x};\mathbf{w}_{\mathrm{back},i},\mathbf{b}_i^{\mathrm{BN}},\xi_i)\in\mathbb{R}^{d_g}$ denote the backbone feature vector of dimension $d_g$, explicitly retaining local normalization dependence. The domain loss $\mathcal{L}_{i,e}^{\mathrm{dom}}
(\mathbf{w}_{\mathrm{back},i},\mathbf{w}_{\mathrm{dom}};\mathbf{b}_i^{\mathrm{BN}},\xi_i)$ is defined as
\begin{equation}\small
-\frac{1}{|\mathcal{B}_{i,e}(t)|}
\sum_{(\mathbf{x},y)\in\mathcal{B}_{i,e}(t)}
\log D_{k(i,t)}
\bigl(g(\mathbf{x};\mathbf{w}_{\mathrm{back},i},\mathbf{b}_i^{\mathrm{BN}},\xi_i);\mathbf{w}_{\mathrm{dom}}\bigr),
\label{eq:domain_loss}
\end{equation}
where $D_k$ is the discriminator probability for zone $k$, with domain label $k(i,t)$ distinct from class label $y$. The discriminator remains fixed at $\mathbf{w}_{\mathrm{dom},i}^{\mathrm{ref}}(t)$ during local training and is updated at the cloud using contributions from multiple zones. The backbone combines supervised learning, a proximal penalty, and the reversed domain gradient through
$
\mathcal{L}_{i,e}^{\mathrm{back}}
=\mathcal{L}_{i,e}^{\mathrm{sup}}
+\frac{\mu}{2}\|\mathbf{w}_{\mathrm{back},i}-\mathbf{w}_{\mathrm{back},i}^{\mathrm{ref}}(t)\|_2^2
-\lambda(t)\mathcal{L}_{i,e}^{\mathrm{dom}},
$
where $\mu\geq0$ controls the proximal penalty. The adversarial weight increases gradually as
$
\lambda(t)=\lambda_{\max}
\min\left\{1,\frac{t}{T_{\mathrm{w}}}\right\},
 T_{\mathrm{w}}=\lceil\eta_{\mathrm{w}}T\rceil,
$
where $\lambda_{\max}\geq0$ is the maximum weight,
$\eta_{\mathrm{w}}\in(0,1]$ is the warm-up fraction, and
$T_{\mathrm{w}}$ is the warm-up duration. A SCAFFOLD-based
correction reduces variation among local backbone updates.
Let $\mathbf{c}_i^t\in\mathbb{R}^{d_{\mathrm{back}}}$ and
$\mathbf{c}_i^{\mathrm{ref}}(t)$ denote the local and global
control variates, respectively. Both are initialized to zero
and fixed during local training. With learning rate
$\eta_l>0$, the updates are
\begin{equation}
\begin{aligned}
&\mathbf{w}_{\mathrm{back},i,e}^{t}
=\mathbf{w}_{\mathrm{back},i,e-1}^{t}-\eta_l
\left(\nabla_{\mathbf{w}_{\mathrm{back},i}}\mathcal{L}_{i,e}^{\mathrm{back}}
-\mathbf{c}_i^t+\mathbf{c}_i^{\mathrm{ref}}(t)\right),\\
&\psi_{i,k,e}^t
=\psi_{i,k,e-1}^t
-\eta_l\nabla_{\psi_{i,k}}\mathcal{L}_{i,e}^{\mathrm{sup}},
\xi_{i,e}^t
=\xi_{i,e-1}^t
-\eta_l\nabla_{\xi_i}\mathcal{L}_{i,e}^{\mathrm{sup}}.
\end{aligned}
\label{eq:local_updates}
\end{equation}
All derivatives are evaluated at the pre-update local state;
components outside the indicated differentiation are held fixed.
The running statistics $\mathbf{b}_i^{\mathrm{BN}}$ are maintained from local
activations. The trained outputs are
$\mathbf{w}_{\mathrm{back},i}^{t}=\mathbf{w}_{\mathrm{back},i,E}^{t}$ and $\psi_{i,k}^t=\psi_{i,k,E}^t$,
while $\xi_i=\xi_{i,E}^t$ is retained locally.
The local control variate is then updated as
$
\mathbf{c}_i^{t+1}
=\operatorname{Clip}\left(
\mathbf{c}_i^t-\mathbf{c}_i^{\mathrm{ref}}(t)
+\frac{\mathbf{w}_{\mathrm{back},i}^{\mathrm{ref}}(t)-\mathbf{w}_{\mathrm{back},i}^{t}}{E\eta_l},
c_{\mathrm{clip}}\right),
$
where $c_{\mathrm{clip}}>0$ is clipping threshold. For
$\mathbf{v}_{\mathrm{clip}}\in\mathbb{R}^{d_{\mathrm{back}}}$, clipping is
\begin{equation}\small
\operatorname{Clip}(\mathbf{v}_{\mathrm{clip}},c_{\mathrm{clip}})
=
\begin{cases}
\mathbf{v}_{\mathrm{clip}},
& \|\mathbf{v}_{\mathrm{clip}}\|_2\leq c_{\mathrm{clip}},\\
c_{\mathrm{clip}}\mathbf{v}_{\mathrm{clip}}/
{\|\mathbf{v}_{\mathrm{clip}}\|_2},
& \|\mathbf{v}_{\mathrm{clip}}\|_2>c_{\mathrm{clip}}.
\end{cases}
\label{eq:control_clipping}
\end{equation}
If training is deferred, the local model and control variate
are retained, with $\mathbf{c}_i^{t+1}=\mathbf{c}_i^t$.

\subsection{Domain-Aware Clustering and CH Selection}
\label{sec:cluster-score}

At reclustering events separated by $K_{\mathrm{cl}}\geq1$ rounds, the framework selects memberships and leaders using environment signatures and adjusted drift risks, with the prescribed number of active clusters $M_t$ fixed during each selection. Let $\mathbf{Z}^t=[z_{i,m}^t]\in\{0,1\}^{N\times M_t}$ denote the leadership indicators and $\mathbf{f}(t)=[f_1(t),\ldots,f_N(t)]^\top$ the frequency vector. The set $\mathcal{R}_t$ contains pairs $(\mathbf{X}^t,\mathbf{Z}^t)$ satisfying \eqref{eq:main_problem} for at least one frequency allocation. Membership, leadership, and risk constraints are checked directly, while resource feasibility follows \eqref{eq:resource_feasibility}. For an existing nonempty cluster, let $\mathcal{C}_m^{t^-}$ denote its membership immediately before reclustering, with reference signature
$
\bar{\mathbf{s}}_m(t)
=\frac{1}{|\mathcal{C}_m^{t^-}|}
\sum_{j\in\mathcal{C}_m^{t^-}}\mathbf{s}_j(t).
$
At initialization or when the prescribed cluster count changes, the first feasible configuration under a fixed lexicographic ordering of binary assignment and leadership entries supplies the reference memberships. Reference signatures remain fixed during the search. For nonzero vectors, their cosine similarity is expressed as
\begin{equation}
\Gamma_{i,m}^{\mathrm{env}}(t)
=\frac{\mathbf{s}_i(t)^\top\bar{\mathbf{s}}_m(t)}
{\|\mathbf{s}_i(t)\|_2\|\bar{\mathbf{s}}_m(t)\|_2},
\label{eq:environment_similarity}
\end{equation}
where $\Gamma_{i,m}^{\mathrm{env}}(t)\in[-1,1]$ measures
signature alignment. If either vector has zero norm, the
similarity is set to zero.
Using the utility in \eqref{eq:cluster_score},
the selected configuration satisfies
$
(\mathbf{X}^{t,\star},\mathbf{Z}^{t,\star})
\in\operatorname*{arg\,max}_{(\mathbf{X}^t,\mathbf{Z}^t)
\in\mathcal{R}_t}\mathcal{Q}_t(\mathbf{X}^t,\mathbf{Z}^t).
$
The finite search enumerates nonempty assignments and eligible
member heads, rejects resource-infeasible candidates, and retains
the highest-scoring pair. Ties follow the same fixed ordering.
This specifies an exact selection procedure without assuming
an efficient solution to the combinatorial search. For the selected configuration, frequencies are assigned as
$f_i(t)=f_{i,\min}(t)$ using
\eqref{eq:minimum_feasible_frequency}. Between reclustering
events, the current configuration is retained only while all
constraints remain satisfied; otherwise, the search is repeated.
If $\mathcal{R}_t$ is empty, training is deferred and monitoring
continues. The score guides network organization and is distinct
from the population learning objective.

\subsection{Hierarchical Aggregation and Synchronization}
\label{sec:global_synchronization}
Each member update includes its round, cluster configuration,
and synchronized source-model version. A \ac{CH} merges only
complete collections with a common round and source version,
and computes the backbone using the weights in
\eqref{eq:aggregation_weights} and the rule in
\eqref{eq:cluster_aggregation}. Let
$\vartheta_i\in\mathbb{R}^{d_\vartheta}$ denote a trained
backbone or active-head parameter block. For a nonempty
labeled sample set $\mathcal{A}_i^t$, its diagonal empirical
Fisher estimate and regularized importance are
$
\widehat{\mathbf{I}}_i^\vartheta
=
\frac{1}{|\mathcal{A}_i^t|}
\sum_{(\mathbf{x},y)\in\mathcal{A}_i^t}
\left[
\nabla_{\vartheta}
\log p_i(y\mid\mathbf{x};\vartheta_i)
\right]^{\odot 2},
\quad
\mathbf{P}_i^\vartheta
=
\widehat{\mathbf{I}}_i^\vartheta
+c_{\mathrm{F}}\mathbf{e}_{\mathrm{ones}},
$
where $p_i$ is the local class probability,
$c_{\mathrm{F}}$ is a strictly positive damping parameter, and
$\mathbf{e}_{\mathrm{ones}}\in\mathbb{R}^{d_\vartheta}$
is the all-ones vector. The estimate averages squared
sample gradients while holding other model components fixed.
For a nonempty contributor set $\mathcal{V}$, the merge is
\begin{equation}\small
\vartheta^{\mathrm{merge}}
=\left(\sum_{i\in\mathcal{V}}\mathbf{P}_i^\vartheta
\right)^{\odot(-1)}
\odot\sum_{i\in\mathcal{V}}
\mathbf{P}_i^\vartheta\odot\vartheta_i.
\label{eq:fisher_merge}
\end{equation}

For each nonempty environment group $\mathcal{C}_{m,k}^t$, the \ac{CH} computes the summed importance vector $\mathbf{A}_{m,k}^t=\sum_{i\in\mathcal{C}_{m,k}^t}\mathbf{P}_i^{\psi_k}$ and importance-weighted parameter sum $\mathbf{B}_{m,k}^t=\sum_{i\in\mathcal{C}_{m,k}^t}\mathbf{P}_i^{\psi_k}\odot\psi_{i,k}^t$. The aggregated head is $\psi_{m,k}^t=(\mathbf{A}_{m,k}^t)^{\odot(-1)}\odot\mathbf{B}_{m,k}^t$. The backbone remains sample-weighted, with cloud importance summary $\mathbf{P}_m^{\mathrm{back},t}=\sum_{i\in\mathcal{C}_m^t}\mathbf{P}_i^{\mathrm{back}}$ evaluated at member models rather than the aggregated backbone. The additive head statistics preserve the direct Fisher merge over member head contributions. The backbone importance summary instead supports a cluster-level Fisher merge of the sample-weighted cluster backbones. Let $s=0,1,\ldots$ index cloud synchronization events, with $s=0$ denoting initialization. At event $s\geq1$, only packets based on version $s-1$ are accepted. A newer round replaces the stored packet for an unchanged configuration, while a newer overlapping configuration replaces older packets sharing its members. Duplicates and packets conflicting with newer stored configurations are rejected, ensuring that each \ac{UAV} appears in at most one accepted packet per event. Let $\mathcal{A}_s$ denote the accepted packet set and $\mathcal{A}_{s,k}$ its subset carrying head $k$. Each packet $q$ contains backbone $\mathbf{w}_{\mathrm{back},q}$, importance vector $\mathbf{P}_q^{\mathrm{back}}$, and head statistics $\mathbf{A}_{q,k},\mathbf{B}_{q,k}$. For nonempty contributing sets, the cloud updates are
$
\mathbf{w}_{\mathrm{back}}^{s}
=
\left(
\sum_{q\in\mathcal{A}_s}\mathbf{P}_q^{\mathrm{back}}
\right)^{\odot(-1)}
\odot
\sum_{q\in\mathcal{A}_s}
\mathbf{P}_q^{\mathrm{back}}\odot\mathbf{w}_{\mathrm{back},q},
$
$
\psi_k^s
=
\left(
\sum_{q\in\mathcal{A}_{s,k}}
\mathbf{A}_{q,k}
\right)^{\odot(-1)}
\odot
\sum_{q\in\mathcal{A}_{s,k}}
\mathbf{B}_{q,k}.
$
An empty contributing set leaves the corresponding backbone or head unchanged. The cloud stores each member's latest accepted control variate $\bar{\mathbf{c}}_i^{\,s}$ and computes $\mathbf{c}^{\,s}=N^{-1}\sum_{i\in\mathcal{N}}\bar{\mathbf{c}}_i^{\,s}$. Older records are rejected, while members without accepted records retain zero vectors.

The sample-weighted domain gradient from local training is
$
\mathbf{d}_{i,\mathrm{dom}}^t
=
\frac{1}{q_i(t)}
\sum_{e=1}^{E}
|\mathcal{B}_{i,e}(t)|
\nabla_{\mathbf{w}_{\mathrm{dom}}}
\mathcal{L}_{i,e}^{\mathrm{dom}},
$
evaluated using pre-update features and the fixed discriminator
reference. Let $\mathcal{P}_s$ contain the members represented
in accepted packets, with $t_i^s$ denoting the originating round
of member $i$. Let $s^{-}$ denote the synchronization event
preceding $s$. All accepted gradients are evaluated using
$\mathbf{w}_{\mathrm{dom}}^{s^{-}}$. If their domain labels
span at least two environments, the cloud updates
\begin{equation}
\mathbf{w}_{\mathrm{dom}}^{s}
=
\mathbf{w}_{\mathrm{dom}}^{s^{-}}
-\eta_d
\frac{
\displaystyle\sum_{i\in\mathcal{P}_s}
q_i(t_i^s)\mathbf{d}_{i,\mathrm{dom}}^{t_i^s}
}{
\displaystyle\sum_{i\in\mathcal{P}_s}
q_i(t_i^s)
},
\label{eq:domain_discriminator_update}
\end{equation}
where $\eta_d>0$ is the learning rate for the accepted domain loss; otherwise, the discriminator remains unchanged. At synchronization, the cloud broadcasts the backbone, heads, discriminator, gate, and global control variate as one versioned collection. Accepted packets are cleared, while control records persist. Payloads $b_i(t)$ and $b_m^{\mathrm{up}}(t)$ contain transmitted parameters, statistics, signatures, sample counts, and provenance metadata. Excluded contributions still incur transmission costs, while $\mathbf{b}_i^{\mathrm{BN}},\xi_i$ remain local.

\subsection{Environment-Gated Inference and Local Adaptation}
\label{sec:gate}

The input signature guides head selection through a softmax gate $G(\mathbf{s};\zeta)$, where $\zeta$ contains its parameters and $G_k$ denotes the probability assigned to zone $k$, with $\sum_{k\in\mathcal{K}}G_k=1$. At synchronization event $s$, let $\mathcal{G}_s$ contain accepted signatures and their zone labels at collection time. For nonempty $\mathcal{G}_s$, the gate-training loss is expressed as
\begin{equation}
\mathcal{L}_{\mathrm{gate}}(\zeta)
=-\frac{1}{|\mathcal{G}_s|}
\sum_{(\mathbf{s},k)\in\mathcal{G}_s}
\log G_k(\mathbf{s};\zeta).
\label{eq:gate_loss}
\end{equation}
The cloud updates $\zeta^s=\zeta^{s-1}-\eta_g\nabla_{\zeta}\mathcal{L}_{\mathrm{gate}}(\zeta^{s-1})$, where $\eta_g>0$ is the learning rate, and retains the previous gate if $\mathcal{G}_s$ is empty. Using the available gate, the $i$-th \ac{UAV} selects
$
\widehat{k}_i(t)=\operatorname*{arg\,max}_{k\in\mathcal{K}}
G_k(\mathbf{s}_i(t);\zeta),
$
with ties resolved by the smallest zone index. Let $\mathbf{p}_{i,k}(\mathbf{x};t)$ denote the class-probability vector produced by its local backbone, head $k$, and normalization components. Inference uses $\mathbf{p}_{i,\widehat{k}_i(t)}$ unless mixing is activated. For $K\geq2$, let $k_{i,1}(t)$ and $k_{i,2}(t)$ index the two largest gate outputs under the same tie rule. Mixing activates when their probability difference is at most $m_{\mathrm{gate}}\in[0,1]$ or the drift indicator defined below equals one. The normalized weights and
mixed prediction are expressed as
\begin{equation}
\begin{aligned}
a_{i,r}^{\mathrm{mix}}(t)
&=\frac{G_{k_{i,r}(t)}(\mathbf{s}_i(t);\zeta)}
{\sum_{v=1}^{2}G_{k_{i,v}(t)}(\mathbf{s}_i(t);\zeta)},\\
\mathbf{p}_i^{\mathrm{mix}}(\mathbf{x};t)
&=\sum_{r=1}^{2}a_{i,r}^{\mathrm{mix}}(t)
\mathbf{p}_{i,k_{i,r}(t)}(\mathbf{x};t),
\end{aligned}
\label{eq:expert_prediction_mixture}
\end{equation}
where $r,v\in\{1,2\}$ index the selected heads
and $a_{i,r}^{\mathrm{mix}}(t)$ is the weight of head
$k_{i,r}(t)$. For $K=1$,
$\mathbf{p}_i^{\mathrm{mix}}$ denotes the single
head's prediction. Drift monitoring uses a nonempty local input batch
$\mathcal{U}_i(t)$. Let
$p_{i,y}^{\mathrm{top}}(\mathbf{x};t)$ denote the
probability of class $y$ from head $\widehat{k}_i(t)$.
Before mixing or adaptation, its batch entropy
is calculated as
\begin{equation}
H_i(t)=-\frac{1}{|\mathcal{U}_i(t)|}
\sum_{\mathbf{x}\in\mathcal{U}_i(t)}
\sum_{y\in\mathcal{Y}}
p_{i,y}^{\mathrm{top}}(\mathbf{x};t)
\log p_{i,y}^{\mathrm{top}}(\mathbf{x};t).
\label{eq:prediction_entropy}
\end{equation}
All entropy expressions use natural logarithms
and $0\log0=0$.
Let $W\geq1$ be the integer monitoring-window
length and $W_t=\min\{W,t+1\}$ the number of
available observations at round $t$.
The smoothed entropy is
$\overline{H}_i(t)=W_t^{-1}
\sum_{r=t-W_t+1}^{t}H_i(r)$,
where $r$ indexes monitoring rounds.
The drift indicator is
$d_i(t)=\mathbb{1}[\overline{H}_i(t)>H_{\mathrm{th}}]$,
where $H_{\mathrm{th}}$ is calibrated using
the same statistic on validation data from
the training environments. This flag indicates
elevated predictive uncertainty and supplies
feedback to \eqref{eq:flagged_risk} in subsequent
rounds. When $d_i(t)=1$, adaptation updates only the
local normalization parameters $\xi_i$.
Let $p_{i,y}^{\mathrm{mix}}(\mathbf{x};t,\xi)$
denote the mixed probability of class $y$
evaluated with candidate parameters $\xi$.
The adaptation loss $\mathcal{L}_i^{\mathrm{adapt}}(\xi;t)$ is defined as
\begin{equation}
\begin{aligned}
-\frac{1}{|\mathcal{U}_i(t)|}
\sum_{\mathbf{x}\in\mathcal{U}_i(t)}
\sum_{y\in\mathcal{Y}}
p_{i,y}^{\mathrm{mix}}(\mathbf{x};t,\xi)\times
\log p_{i,y}^{\mathrm{mix}}(\mathbf{x};t,\xi).
\end{aligned}
\label{eq:adaptation_loss}
\end{equation}
Starting from the retained $\xi_i^{(0)}=\xi_i$,
adaptation applies
$\xi_i^{(a+1)}=\xi_i^{(a)}-\eta_{\mathrm{tta}}
\nabla_{\xi}\mathcal{L}_i^{\mathrm{adapt}}
(\xi_i^{(a)};t)$,
where $a=0,\ldots,A_{\mathrm{tta}}-1$ indexes
the $A_{\mathrm{tta}}\geq1$ adaptation steps
and $\eta_{\mathrm{tta}}>0$ is the learning rate.
The backbone, heads, running statistics, gate,
and mixing weights remain fixed.
Adaptation uses unlabeled inputs, and the final
normalization parameters are retained for
subsequent inference and training.

\section{Performance Analysis}
\label{sec:conv}
This section analyzes backbone stationarity, Fisher-weighted
merging, and the communication and computation requirements
of \ac{DA-HFL}.

\subsection{Stationarity of the Aggregated Backbone}
\label{sec:conditional_descent}

At synchronization event $s$, all accepted packets originate
from backbone $\mathbf{w}_{\mathrm{back}}^{s-1}$. For packet $q\in\mathcal{A}_s$,
let $\mathcal{C}_q$ denote its member set and $w_{i,q}$ the
sample-count weight defined in \eqref{eq:aggregation_weights}.
The average corrected direction for member $i$ is
$\bar{\mathbf{u}}_{i,q}
=E^{-1}\sum_{e=1}^{E}\mathbf{u}_{i,e}^{q}$, where
$\mathbf{u}_{i,e}^{q}$ is the direction in
\eqref{eq:local_updates}. Its trained backbone is therefore
$\mathbf{w}_{\mathrm{back}}^{s-1}-\eta_{\mathrm{eff}}\bar{\mathbf{u}}_{i,q}$,
where $\eta_{\mathrm{eff}}=E\eta_l$.
For a nonempty accepted packet set, the cloud assigns each
packet the diagonal weight matrix
\begin{equation} \small
\mathbf{W}_{q,s}
=
\operatorname{diag}\left(
\mathbf{P}_q^{\mathrm{back}}\odot
\left[
\sum_{r\in\mathcal{A}_s}
\mathbf{P}_r^{\mathrm{back}}
\right]^{\odot(-1)}
\right).
\label{eq:cloud_weight_matrix}
\end{equation}
Positive Fisher damping ensures positive denominators and
$\sum_{q\in\mathcal{A}_s}\mathbf{W}_{q,s}
=\mathbf{I}_{d_{\mathrm{back}}}$. Combining the cluster and cloud
aggregation rules gives
\begin{equation}
\begin{aligned}
\mathbf{G}_s
&=\sum_{q\in\mathcal{A}_s}\mathbf{W}_{q,s}
\sum_{i\in\mathcal{C}_q}w_{i,q}\bar{\mathbf{u}}_{i,q},
&\mathbf{w}_{\mathrm{back}}^{s}&=\mathbf{w}_{\mathrm{back}}^{s-1}-\eta_{\mathrm{eff}}\mathbf{G}_s,
\end{aligned}
\label{eq:effective_backbone_update}
\end{equation}
where $\mathbf{G}_s$ denotes the aggregated update direction.
If no packet is accepted, $\mathbf{G}_s=\mathbf{0}$ and
the backbone remains unchanged. The learning objective with heads and normalization states
fixed at event $s-1$ is
$\mathcal{J}_s(\mathbf{w}_{\mathrm{back}})
=\mathcal{J}\bigl(
\mathbf{w}_{\mathrm{back}},\boldsymbol{\psi}^{s-1};
\mathbf{b}^{\mathrm{BN},s-1},\boldsymbol{\xi}^{s-1}\bigr).$
The normalization states remain local; their appearance here
does not require transmission. Assume that each
$\mathcal{J}_s$ is differentiable and has an $L$-Lipschitz
backbone gradient over the analyzed updates, with $L>0$.
The difference between the aggregated direction and the
objective gradient is
$\mathbf{r}_s=\mathbf{G}_s
-\nabla\mathcal{J}_s(\mathbf{w}_{\mathrm{back}}^{s-1})$.
This residual includes the effects of local training,
regularization, control correction, participation, and
Fisher weighting, and is not assumed to have zero mean.
By $L$-smoothness and \eqref{eq:effective_backbone_update},
the backbone update satisfies
\begin{equation}
\begin{aligned}
\mathcal{J}_s(\mathbf{w}_{\mathrm{back}}^{s})
\leq{}&\mathcal{J}_s(\mathbf{w}_{\mathrm{back}}^{s-1})
-\frac{\eta_{\mathrm{eff}}}{2}
\|\nabla\mathcal{J}_s(\mathbf{w}_{\mathrm{back}}^{s-1})\|_2^2
+\frac{\eta_{\mathrm{eff}}}{2}\|\mathbf{r}_s\|_2^2,
\end{aligned}
\label{eq:conditional_descent}
\end{equation}
for $0<\eta_{\mathrm{eff}}\leq1/L$.
This follows from
$2\langle\mathbf{a},\mathbf{b}\rangle
=\|\mathbf{a}\|_2^2+\|\mathbf{b}\|_2^2
-\|\mathbf{a}-\mathbf{b}\|_2^2$
and omission of the nonpositive term
$-\eta_{\mathrm{eff}}(1-L\eta_{\mathrm{eff}})
\|\mathbf{G}_s\|_2^2/2$. Changes in heads and normalization states introduce the
objective variation
$
\Delta_s=
\max\!\left\{
0,\,
\mathcal{J}_{s+1}(\mathbf{w}_{\mathrm{back}}^{s})
-\mathcal{J}_{s}(\mathbf{w}_{\mathrm{back}}^{s})
\right\}.
$
For $S_{\mathrm{sync}}\geq1$ synchronization events
and a common lower bound $\mathcal{J}_{\mathrm{lb}}$,
summing \eqref{eq:conditional_descent} and accounting
for $\Delta_s$ yields

\begin{equation} \small
\begin{split}
\frac{1}{S_{\mathrm{sync}}}
\sum_{s=1}^{S_{\mathrm{sync}}}
\|\nabla\mathcal{J}_s(\mathbf{w}_{\mathrm{back}}^{s-1})\|_2^2
\leq
\frac{2[\mathcal{J}_1(\mathbf{w}_{\mathrm{back}}^o)-\mathcal{J}_{\mathrm{lb}}]}
{\eta_{\mathrm{eff}}S_{\mathrm{sync}}}\\
{}+
\frac{1}{S_{\mathrm{sync}}}
\sum_{s=1}^{S_{\mathrm{sync}}}\|\mathbf{r}_s\|_2^2
+
\frac{2}{\eta_{\mathrm{eff}}S_{\mathrm{sync}}}
\sum_{s=1}^{S_{\mathrm{sync}}-1}\Delta_s .
\end{split}
\label{eq:conv_bound}
\end{equation}
For a fixed step size, the initial-gap term vanishes as $S_{\mathrm{sync}}$ increases. Average stationarity follows if the average squared residual and the average positive objective variation vanish; the protocol alone does not guarantee these conditions.

\subsection{Optimality of Fisher-Weighted Merging}
\label{sec:fisher_analysis}

For a nonempty contributor set $\mathcal{V}$, let
$\mathbf{H}_i=\operatorname{diag}(\mathbf{P}_i^\vartheta)$
denote the regularized importance matrix for parameter
block $\vartheta$. The importance-weighted quadratic
discrepancy is
\begin{equation}
\mathcal{Q}_{\mathrm{F}}(\vartheta)
=\frac{1}{2}\sum_{i\in\mathcal{V}}
(\vartheta-\vartheta_i)^\top
\mathbf{H}_i(\vartheta-\vartheta_i).
\label{eq:fisher_quadratic}
\end{equation}
Here, $\mathbf{H}_i$ represents empirical parameter importance,
not the loss Hessian. Since $c_{\mathrm{F}}>0$,
the Hessian of $\mathcal{Q}_{\mathrm{F}}$,
$\sum_{i\in\mathcal{V}}\mathbf{H}_i$, is positive definite.
Setting the gradient to zero therefore gives the unique
minimizer in \eqref{eq:fisher_merge}. Expanding around this
minimizer yields, for any comparison block
$\vartheta^{\mathrm{cmp}}$ of the same dimension,
\begin{equation}\small
\begin{aligned}
&\mathcal{Q}_{\mathrm{F}}(\vartheta^{\mathrm{cmp}})
-\mathcal{Q}_{\mathrm{F}}(\vartheta^{\mathrm{merge}})\\
&\quad=\frac{1}{2}
(\vartheta^{\mathrm{cmp}}-\vartheta^{\mathrm{merge}})^\top
\left(\sum_{i\in\mathcal{V}}\mathbf{H}_i\right)
(\vartheta^{\mathrm{cmp}}-\vartheta^{\mathrm{merge}})
\geq0.
\end{aligned}
\label{eq:fisher_bound}
\end{equation}
Thus, Fisher merging attains no greater quadratic discrepancy
than a sample-weighted average, with equality exactly when
the two merged blocks coincide. For environment heads, the additive statistics preserve
this result over accepted member contributions. For the
cloud backbone, it applies to cluster backbones and their
importance summaries, since within-cluster averaging remains
sample-weighted. This optimality concerns only
$\mathcal{Q}_{\mathrm{F}}$ and does not establish lower
population loss or higher test accuracy.
\begin{figure*}[t]
\centering
\subfloat[]{%
\includegraphics[width=0.3\textwidth]{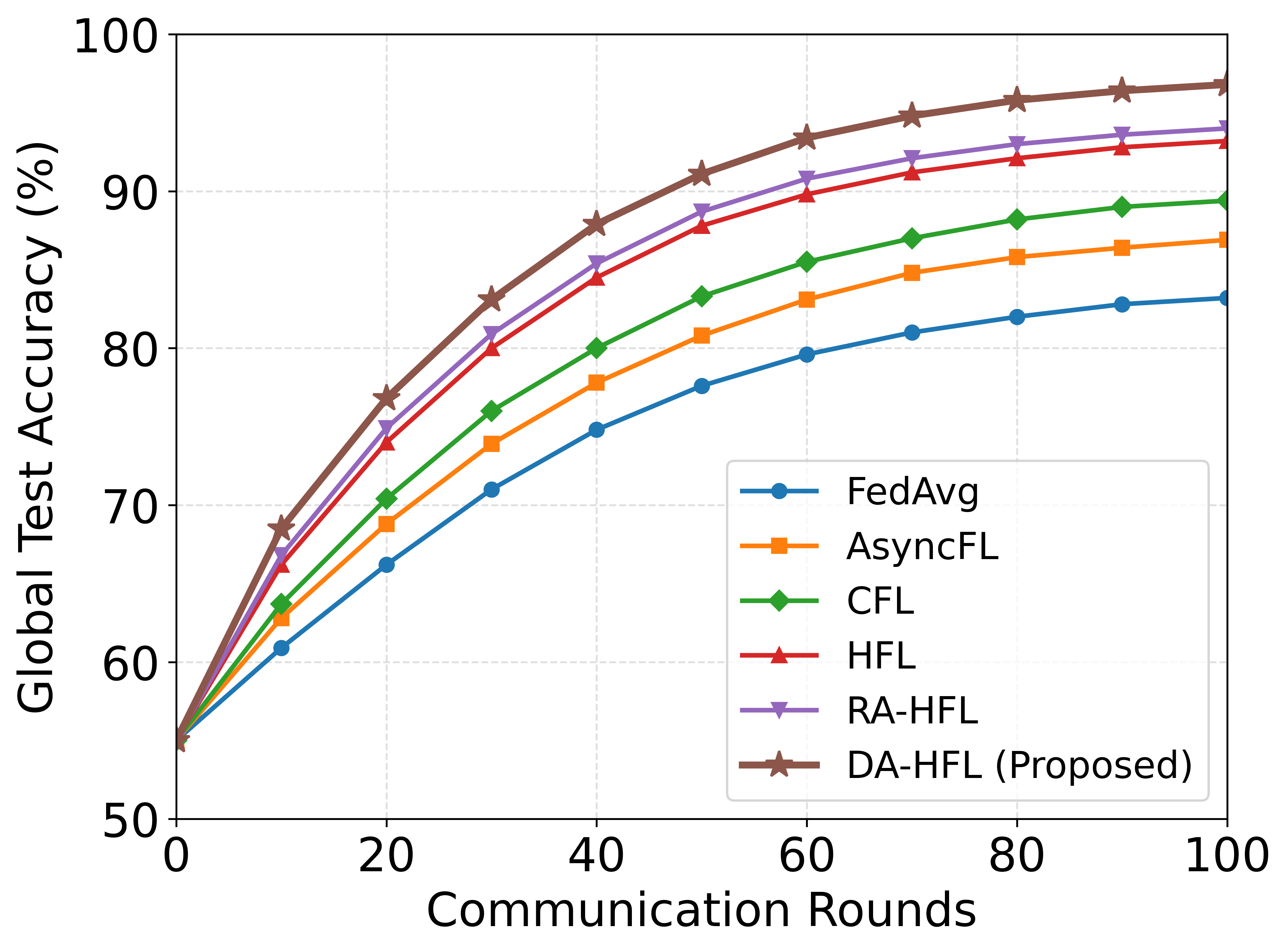}}
\hfil
\subfloat[]{%
\includegraphics[width=0.3\textwidth]{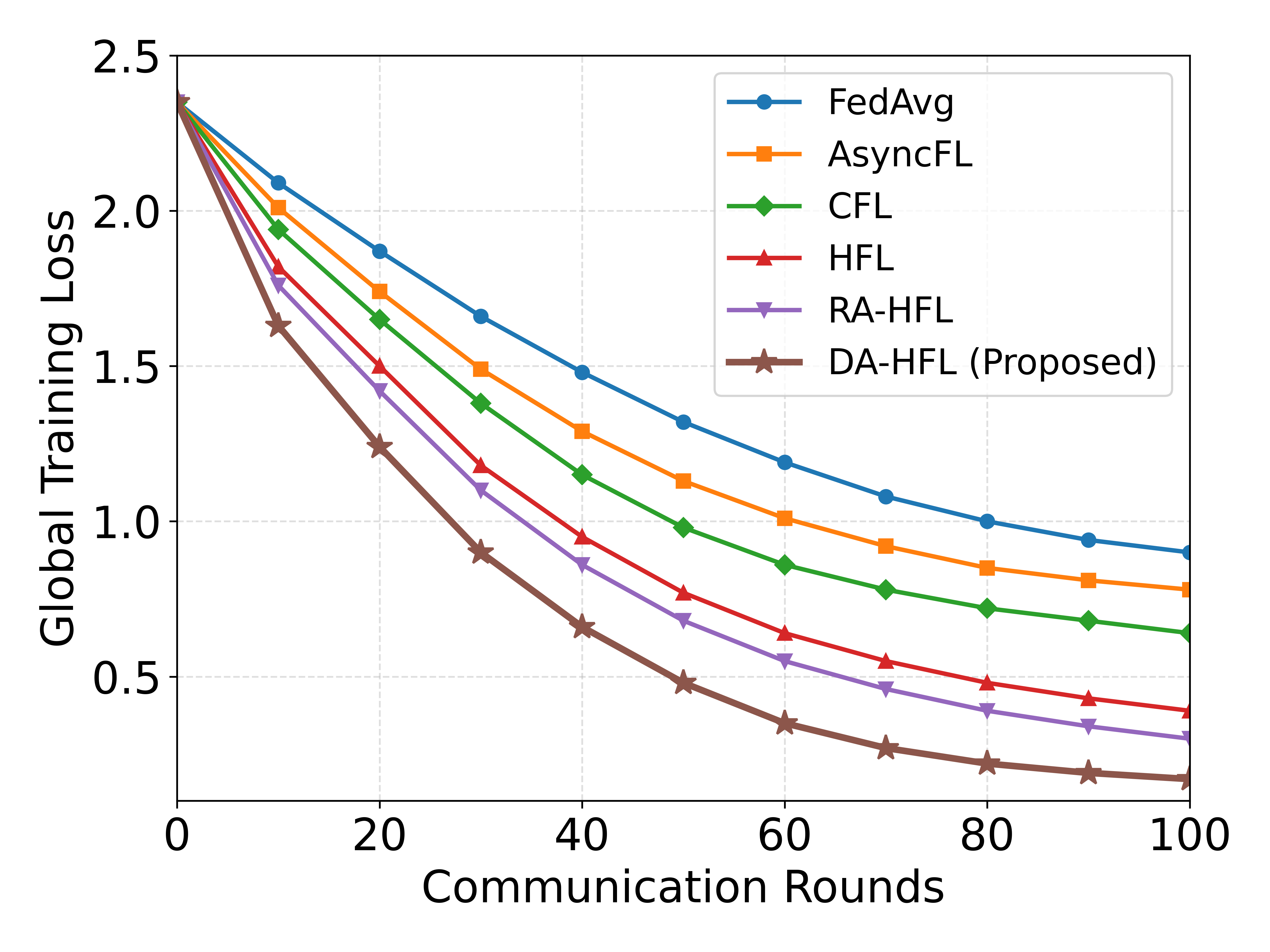}}
\hfil
\subfloat[]{%
\includegraphics[width=0.3\textwidth]{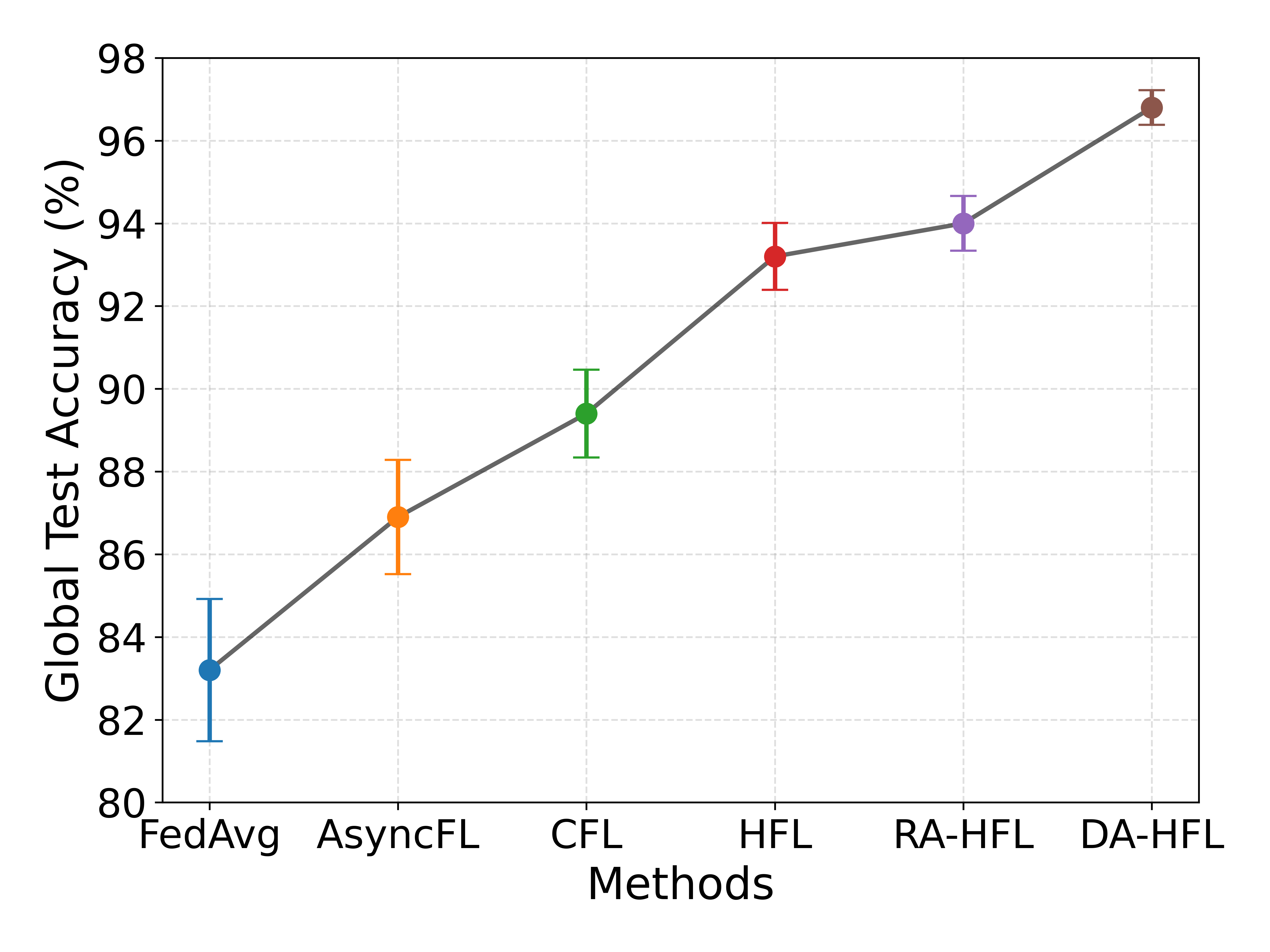}}
\caption{(a) global test accuracy vs.\ communication rounds,
(b) global training loss vs.\ communication rounds and (c) accuracy with temporal error bars.}
\label{fig:convergence}
\end{figure*}
\subsection{Communication and Computation Requirements}
\label{sec:complexity}
Communication is counted in fixed-precision numerical values,
excluding metadata, per cluster uplink and per receiver downlink.
The analytical single-model FedAvg-HFL reference exchanges
$O(d_{\mathrm{back}}+d_\psi)$ values in either direction; naive multi-head
HFL exchanges $O(d_{\mathrm{back}}+Kd_\psi)$ and selects a head using
a supplied zone label. For \ac{DA-HFL}, uplink and downlink costs
are $O(n_m^t(d_{\mathrm{back}}+d_{\mathrm{dom}}+d_x)+K_m(t)d_\psi)$ and
$O(d_{\mathrm{back}}+Kd_\psi+d_{\mathrm{dom}}+d_\zeta)$, respectively, where
$d_{\mathrm{dom}},d_\zeta$ are discriminator and gate parameter counts,
and $K_m(t)=|\{k\in\mathcal{K}:\mathcal{C}_{m,k}^t
\neq\emptyset\}|\leq\min\{K,n_m^t\}$.
The member-to-head payload is $O(d_{\mathrm{back}}+d_\psi+d_{\mathrm{dom}}+d_x)$.
These are per-exchange costs; dividing the \ac{DA-HFL}
downlink cost by $K_{\mathrm{sync}}$ gives its amortized
per-round cost. Actual transmissions include metadata and
incur costs even when contributions are discarded. Let $\mathsf{C}_{\mathrm{back}}$, $\mathsf{C}_{\mathrm{head}}$,
and $\mathsf{C}_{\mathrm{gate}}$ denote forward costs.
Both references require
$O(\mathsf{C}_{\mathrm{back}}+\mathsf{C}_{\mathrm{head}})$
for inference, while \ac{DA-HFL} mixed inference requires
$O(\mathsf{C}_{\mathrm{back}}+2\mathsf{C}_{\mathrm{head}}
+\mathsf{C}_{\mathrm{gate}}+K)$ for a precomputed signature.
The $K$ term accounts for top-two head selection. Local training and Fisher estimation require
$O(q_i(t)\mathsf{C}_{\mathrm{train}}+Ed_{\mathrm{back}})$ and
$O(|\mathcal{A}_i^t|\mathsf{C}_{\mathrm{F}})$ operations,
where $\mathsf{C}_{\mathrm{train}}$ and $\mathsf{C}_{\mathrm{F}}$
are their per-sample gradient costs.
Signature calculation and cluster aggregation cost
$O(|\mathcal{B}_i(t)|d_x)$ and $O(n_m^t(d_{\mathrm{back}}+d_\psi))$.
Cloud merging and auxiliary updates require
$O(|\mathcal{A}_s|d_{\mathrm{back}}+
\sum_{q\in\mathcal{A}_s}K_qd_\psi+
|\mathcal{P}_s|d_{\mathrm{dom}}+Nd_{\mathrm{back}})$,
where $K_q$ counts heads in packet $q$.
Gate training adds
$O(|\mathcal{G}_s|\mathsf{C}_{\mathrm{gate,grad}})$,
with per-signature gradient cost
$\mathsf{C}_{\mathrm{gate,grad}}$.
Activated adaptation costs
$O(A_{\mathrm{tta}}|\mathcal{U}_i(t)|
\mathsf{C}_{\mathrm{adapt}})$,
where $\mathsf{C}_{\mathrm{adapt}}$ is the per-sample
forward and backward cost. The energy-model exclusions
remain unchanged.
Exact clustering considers at most $M_t^N$ labeled assignments
and $N^{M_t}$ head selections per assignment.
With precomputed similarities and fixed-dimensional link
calculations, each candidate costs $O(NM_t)$ to score and test,
giving the conservative bound
$
O\bigl(M_t^N N^{M_t}NM_t\bigr),
$
plus $O(NM_td_x)$ similarity computation.
This exhaustive search does not establish real-time
scalability for large swarms.

\section{Simulation and Results}
\label{sec:simulation}
\label{sec:testbed}
The evaluation considers a feature-level simulation with
$N=40$ \acp{UAV} operating across four environmental
zones, namely city, village, forest, and lake.
The classification task comprises five classes:
person, vehicle, building, vegetation, and water.
Each zone distribution is generated from a shared 24-dimensional class prototype that is shifted by a fixed zone-specific offset and corrupted by isotropic Gaussian noise of variance $0.05$; mobility-induced transitions simply replace the active offset at the prescribed mission times, thereby producing controlled domain shifts without requiring a real \ac{UAV} imagery dataset. Environmental
ambiguity is introduced by moving selected class
prototypes toward a common midpoint. The affected
pairs are vehicle/building in city, building/vegetation
in village, person/vegetation in forest, and
person/water in lake. During the mission, 12 \acp{UAV} change zones at
staggered times, while the remaining \acp{UAV} retain
their initial zones. Resource profiles are sampled independently of zone
identity. Each zone provides 400 held-out evaluation
samples that are disjoint from the training data.
The experiment uses synthetic input features and
therefore evaluates classification rather than
image-based object detection or localization. The reported training configuration uses $T=100$
rounds, $E=5$ local steps per round, a batch size
of 32, and a local learning rate of
$\eta_l=10^{-3}$. Reclustering and synchronization
intervals are both five rounds, with
$K_{\mathrm{cl}}=K_{\mathrm{sync}}=5$. The simulation uses $M_t=4$
clusters, with a mean configuration-search runtime of $0.18$ seconds per reclustering event.
The risk threshold and feedback increment are
$r_{\mathrm{thresh}}=0.3$ and $a_{\mathrm{risk}}=0.2$,
respectively. Entropy monitoring uses a five-round
window, and the reported threshold is the 95th
percentile of validation entropy. The adversarial
weight reaches $\lambda_{\max}=1$ after a warm-up
fraction of $\eta_{\mathrm{w}}=0.2$, corresponding
to the first 20 rounds. The proximal coefficient
is $\mu=0.01$, Fisher estimation uses 10 batches,
and the local adaptation learning rate is
$\eta_{\mathrm{tta}}=10^{-4}$.

The comparison includes FedAvg
\cite{mcmahan2017communication}, which maintains
one synchronous global model; AsyncFL
\cite{10452821,wu2020safa}, which uses
staleness-weighted asynchronous updates; and a
CFL variant \cite{sattler2021clustered} with
fixed initial-environment assignments.
The hierarchical baselines are \ac{HFL}
\cite{liu2020client}, with resource-aware
reclustering, and RA-HFL
\cite{he2023threestage,wang2024uav}, with
resource-aware organization and event-driven
synchronization. The comparison uses a common
swarm, data-generation scenario, local-step
budget, and batch size. These baselines provide
comparisons with global learning, asynchronous
learning, fixed clustering, and resource-aware
hierarchical organization.
\begin{figure}
\centering
\subfloat[]{
  \includegraphics[width=41.9mm,height=33mm]{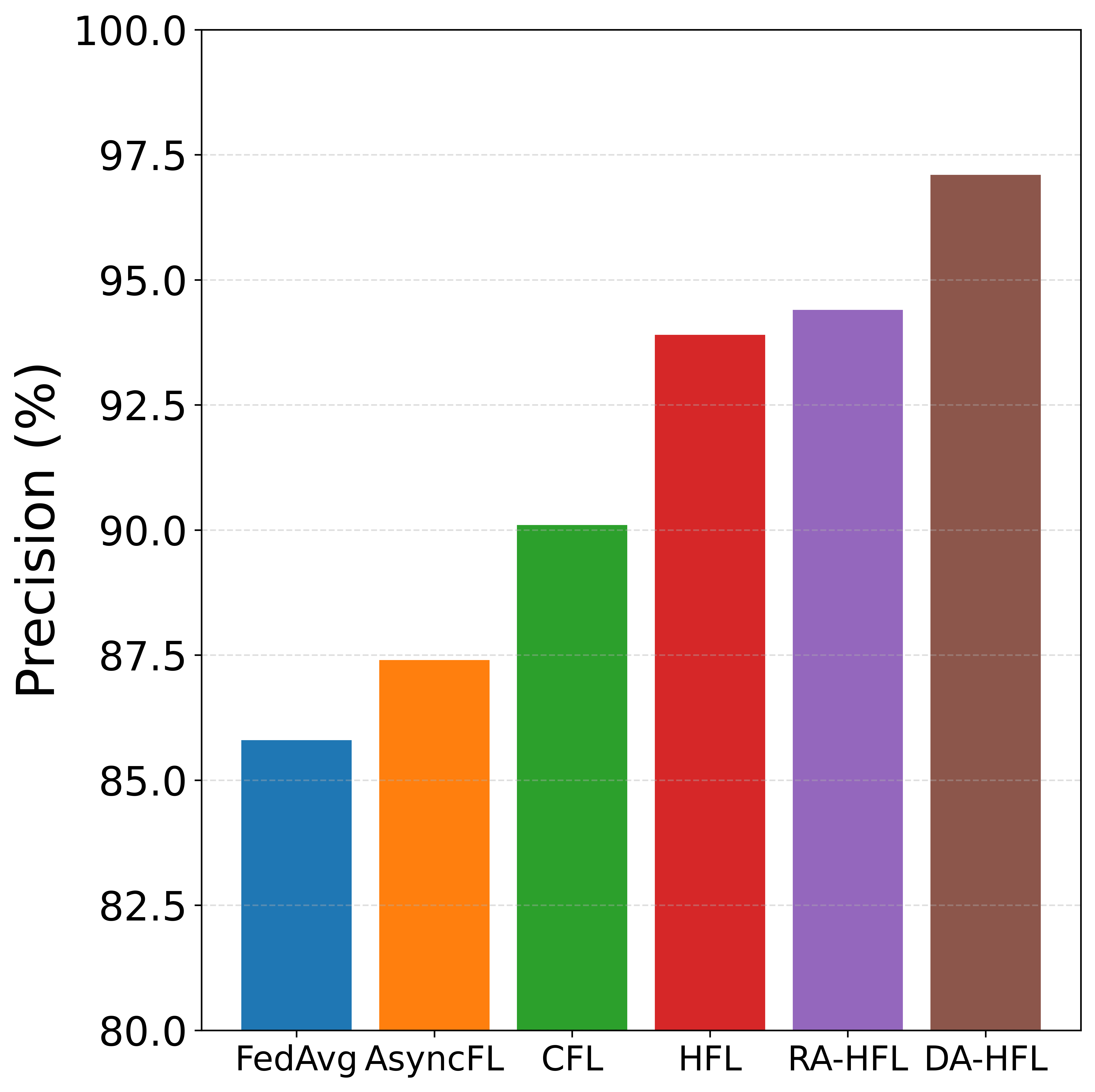}
}
\subfloat[]{
  \includegraphics[width=41.9mm,height=33mm]{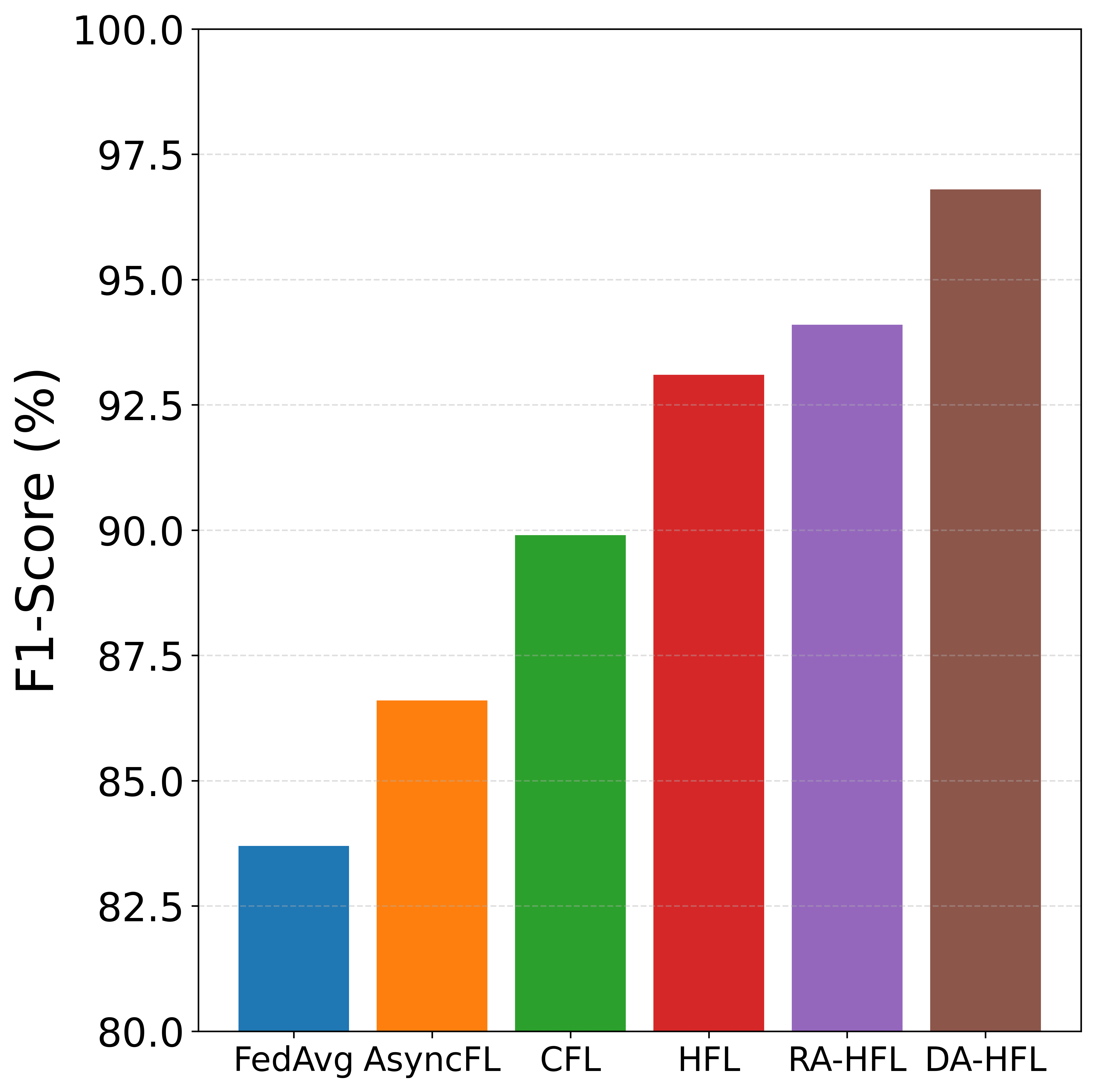}
  
}

\subfloat[]{
  \includegraphics[width=41.9mm,height=33mm]{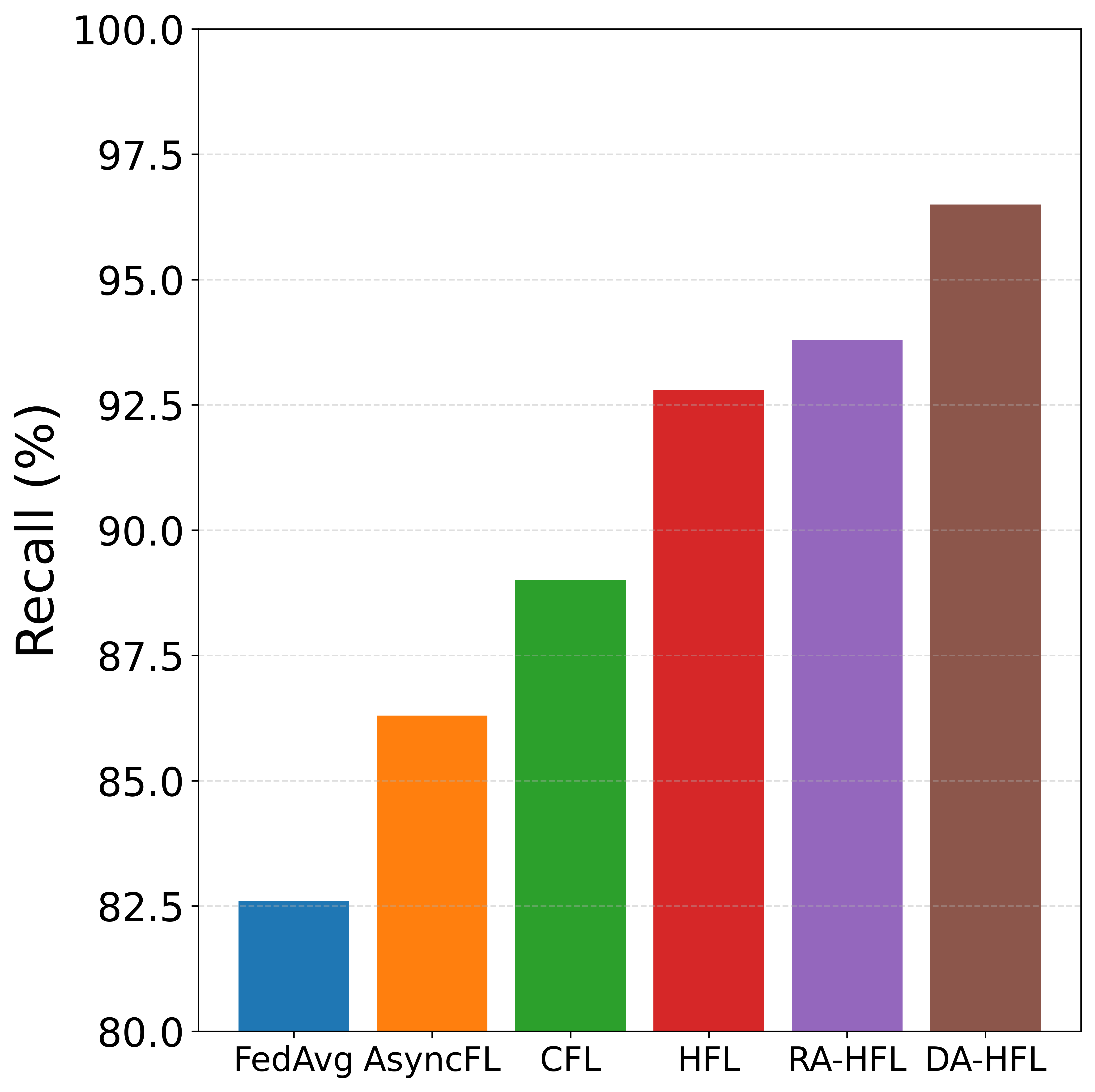}
}
\subfloat[]{
  \includegraphics[width=41.9mm,height=33mm]{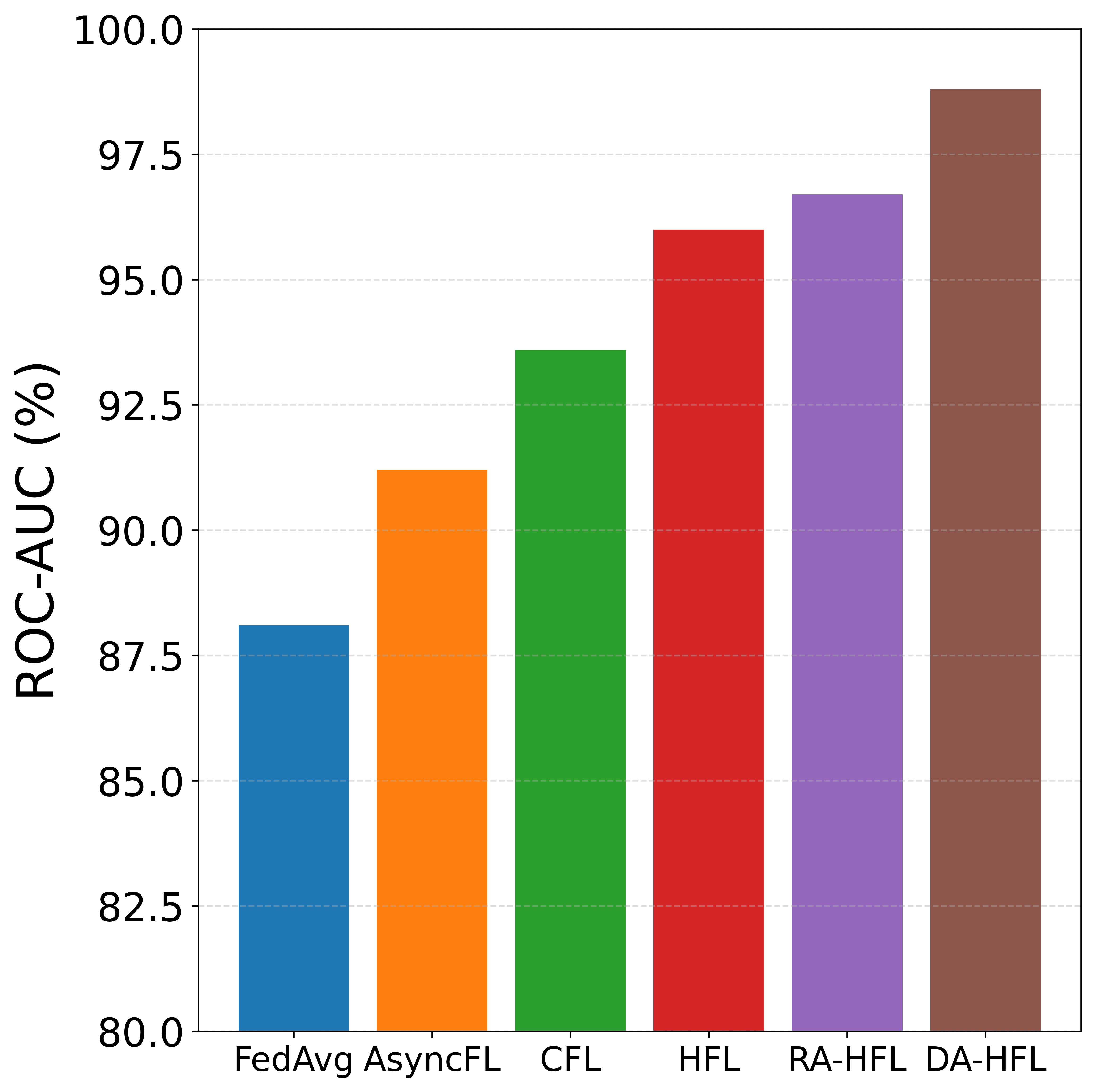}
}
\caption{Classification performance: (a) precision,
(b) F1-score, (c) recall, and (d) ROC-AUC.}
\label{fig:performance_comparison}
\end{figure}
\begin{figure*}
\centering
\includegraphics[width=0.8\textwidth]{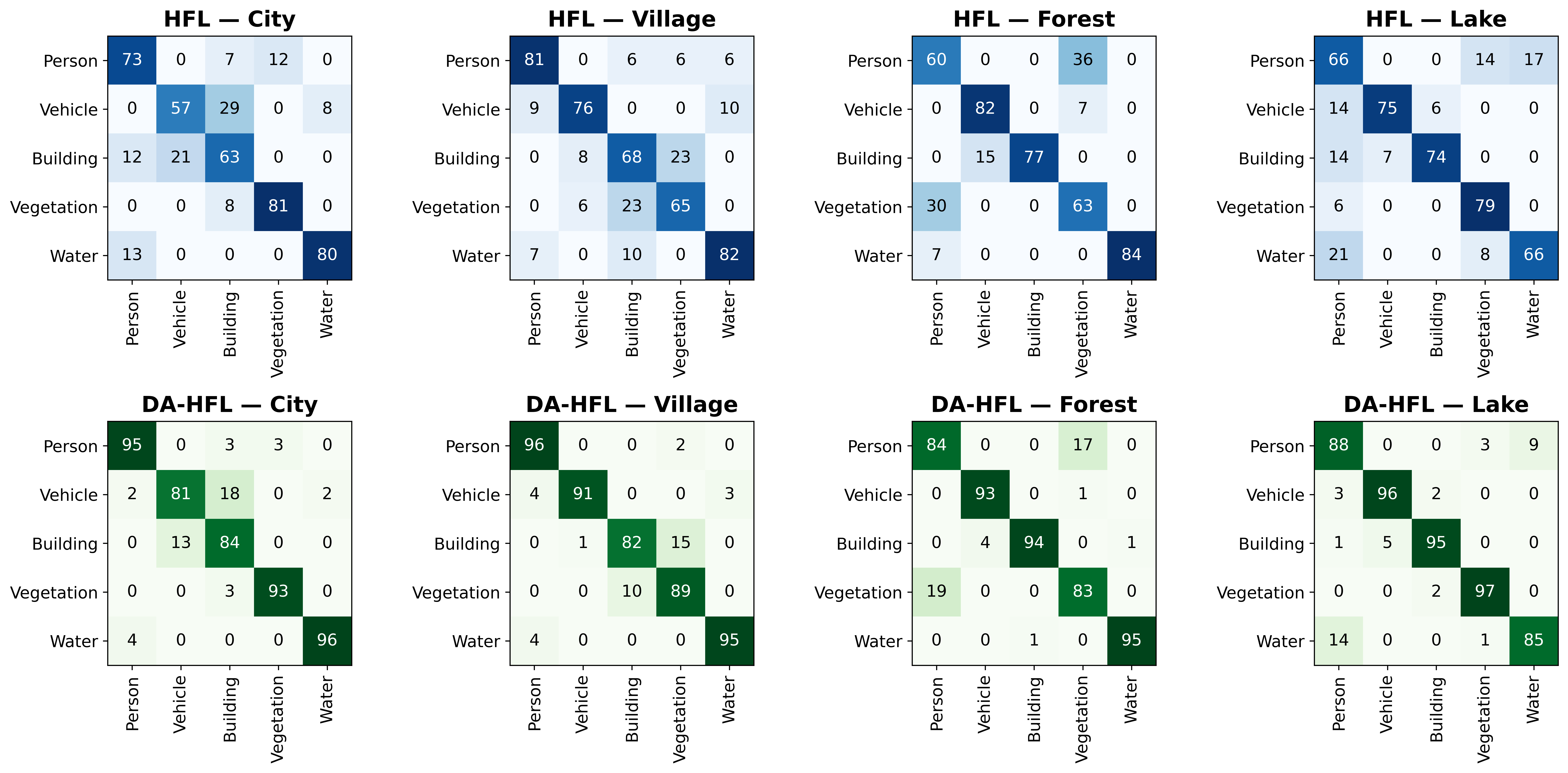}
\caption{Displayed confusion matrices for \ac{HFL}
(top row) and \ac{DA-HFL} (bottom row) across the four zones.}
\label{fig:matrices}
\end{figure*}

\begin{figure}
\centering
\subfloat[]{
  \includegraphics[width=41.9mm,height=36mm]{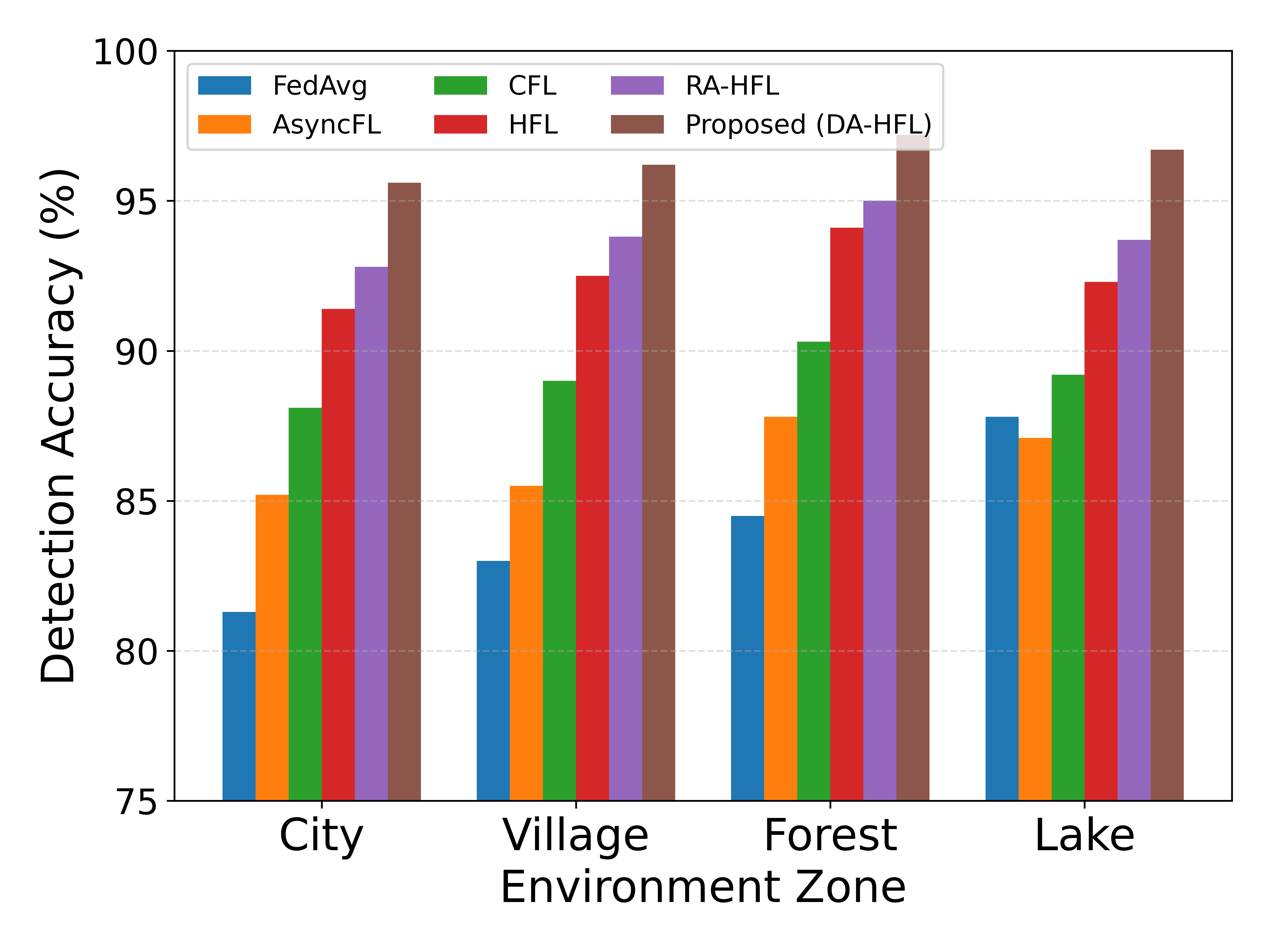}
}
\subfloat[]{
  \includegraphics[width=41.9mm,height=36mm]{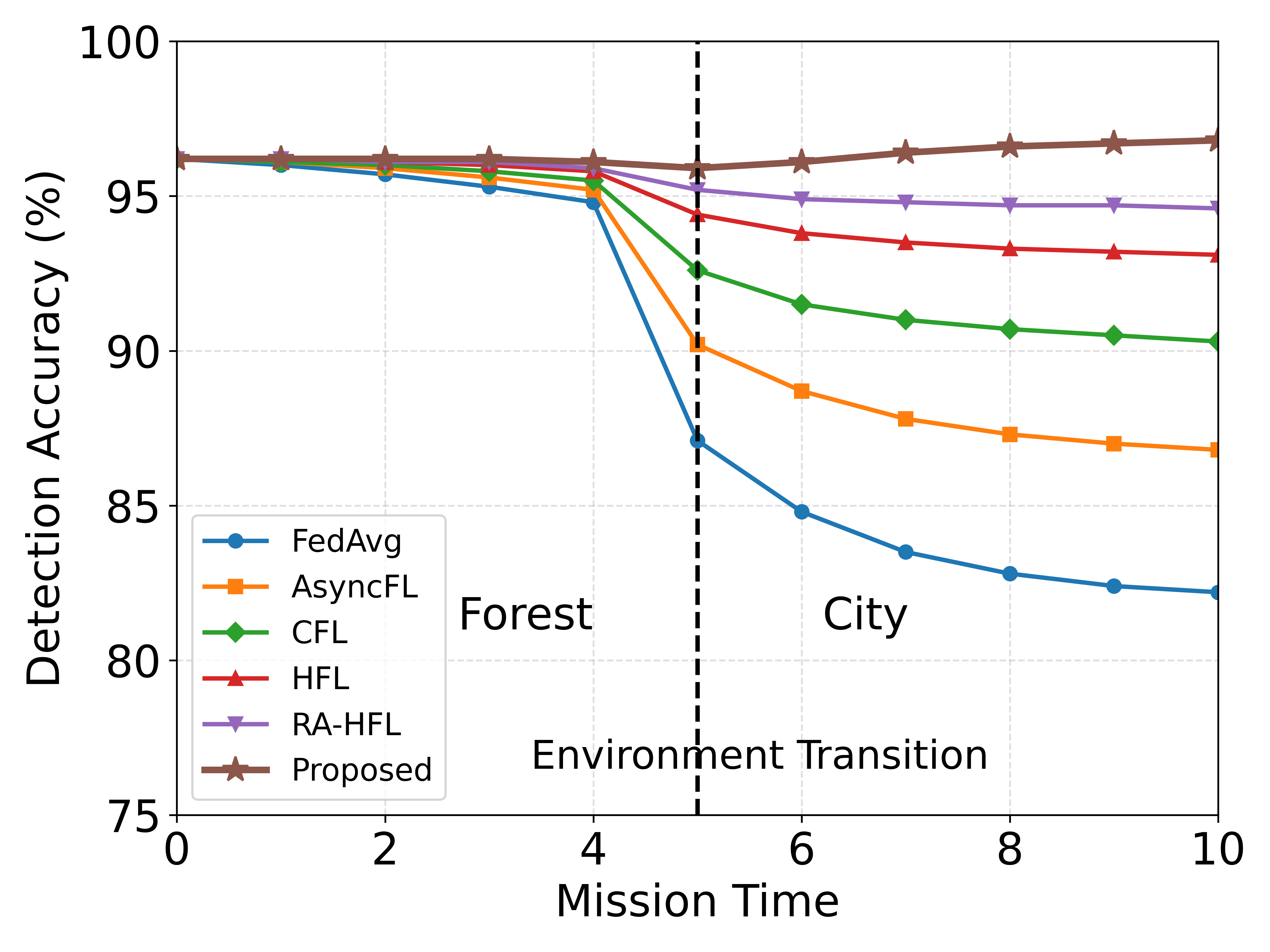}
  
}

\subfloat[]{
  \includegraphics[width=41.9mm,height=36mm]{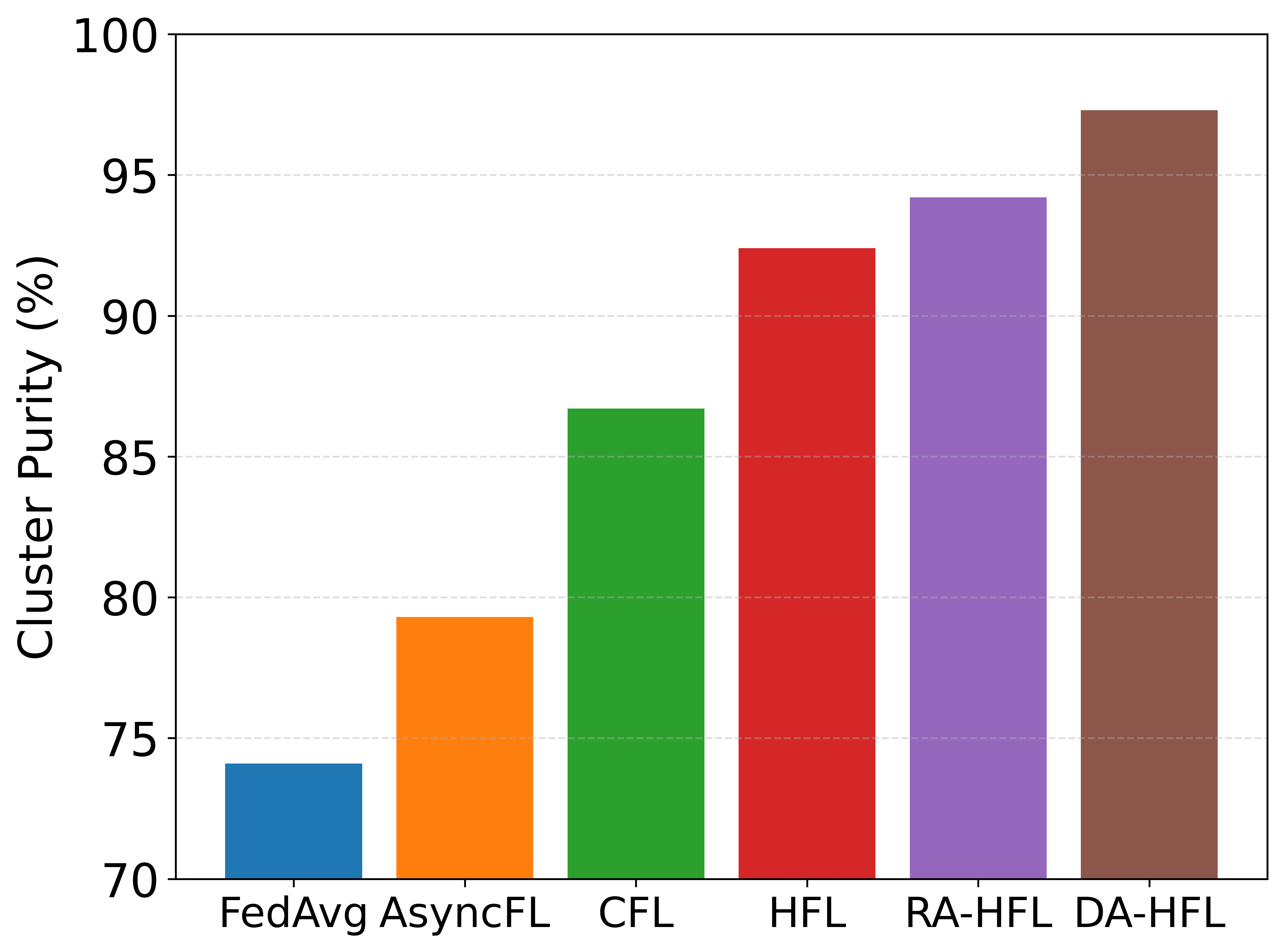}
}
\subfloat[]{
  \includegraphics[width=41.9mm,height=36mm]{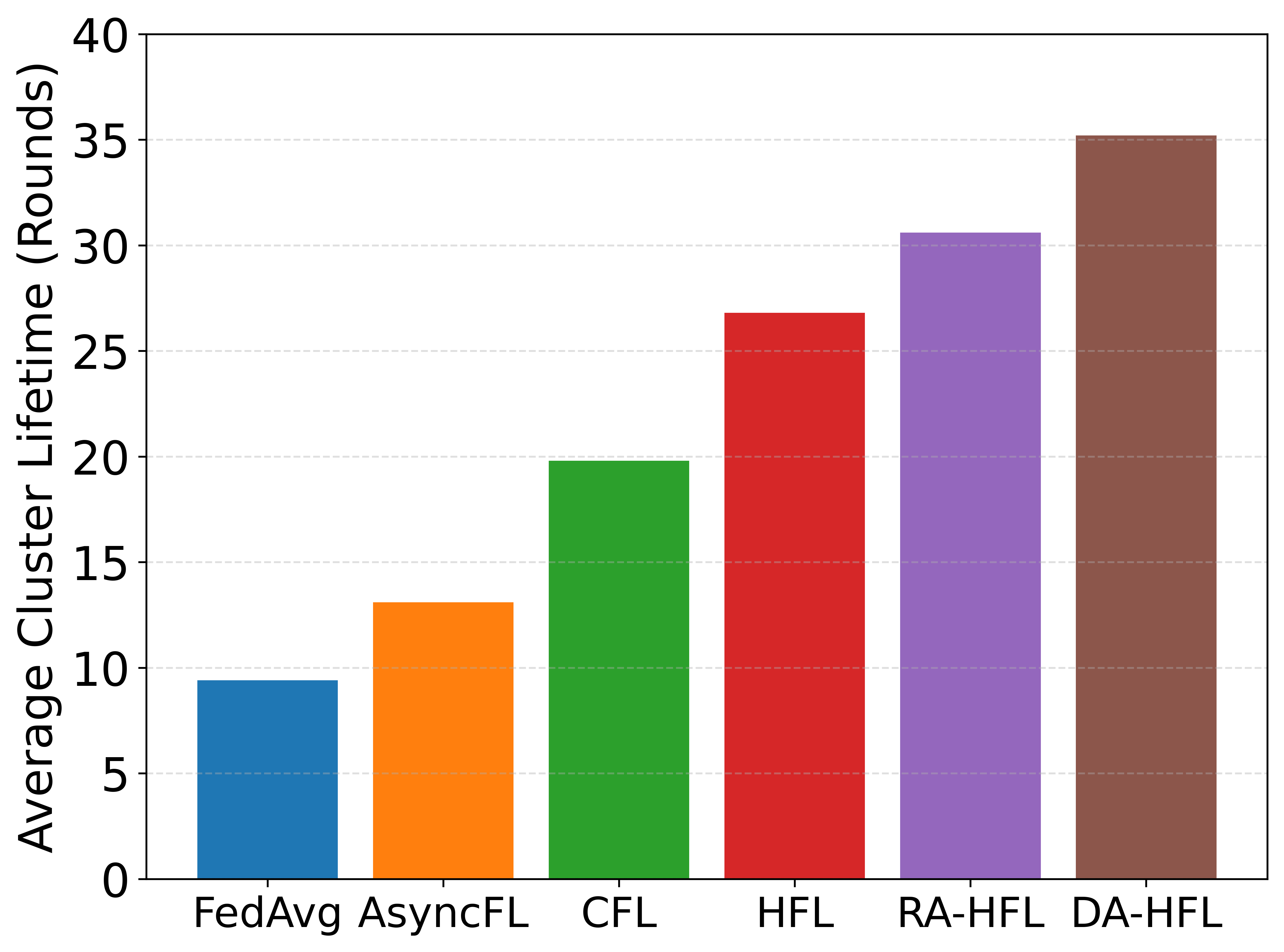}
}

\subfloat[]{
  \includegraphics[width=41.9mm,height=36mm]{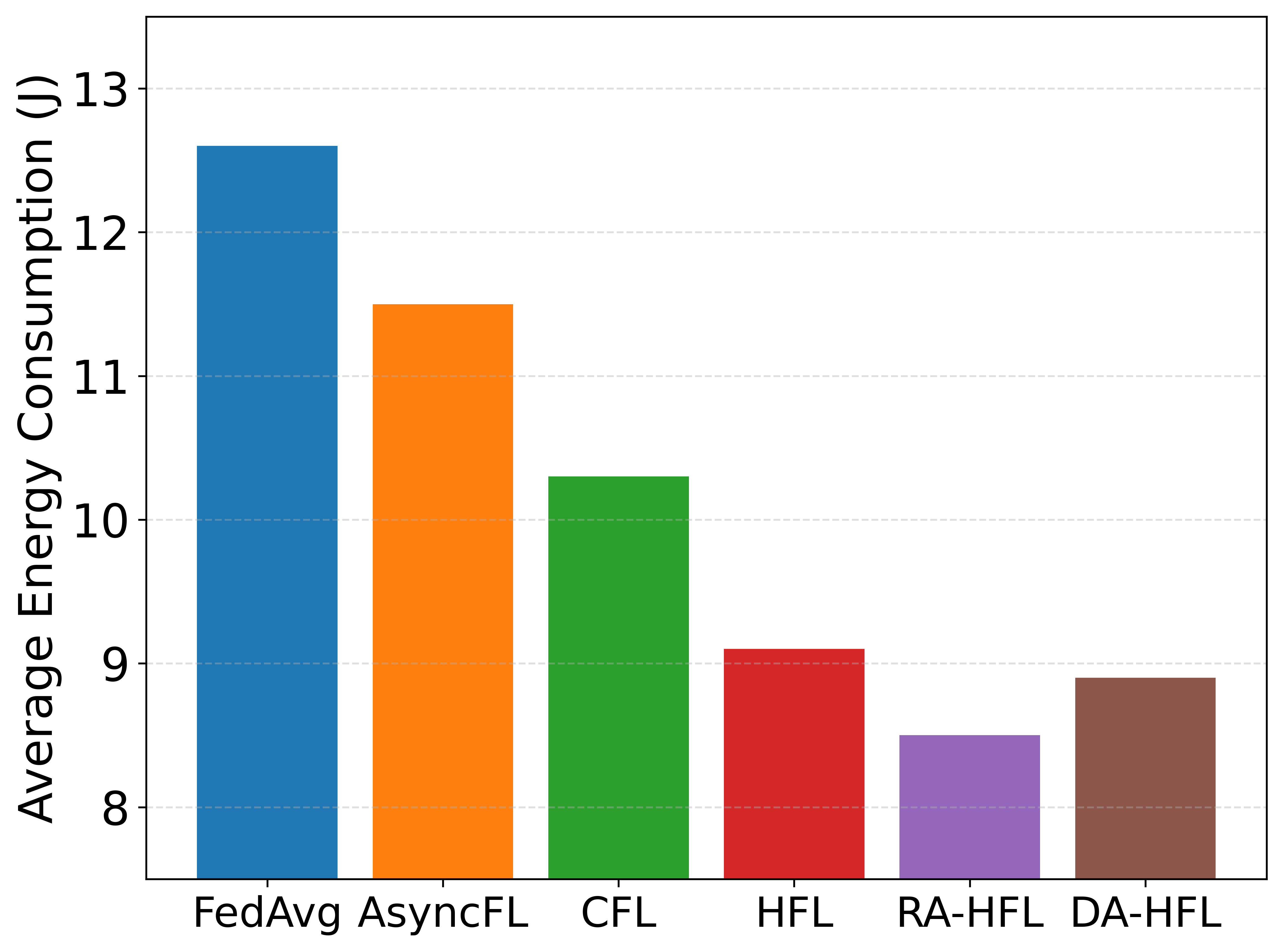}
}
\subfloat[]{
  \includegraphics[width=41.9mm,height=36mm]{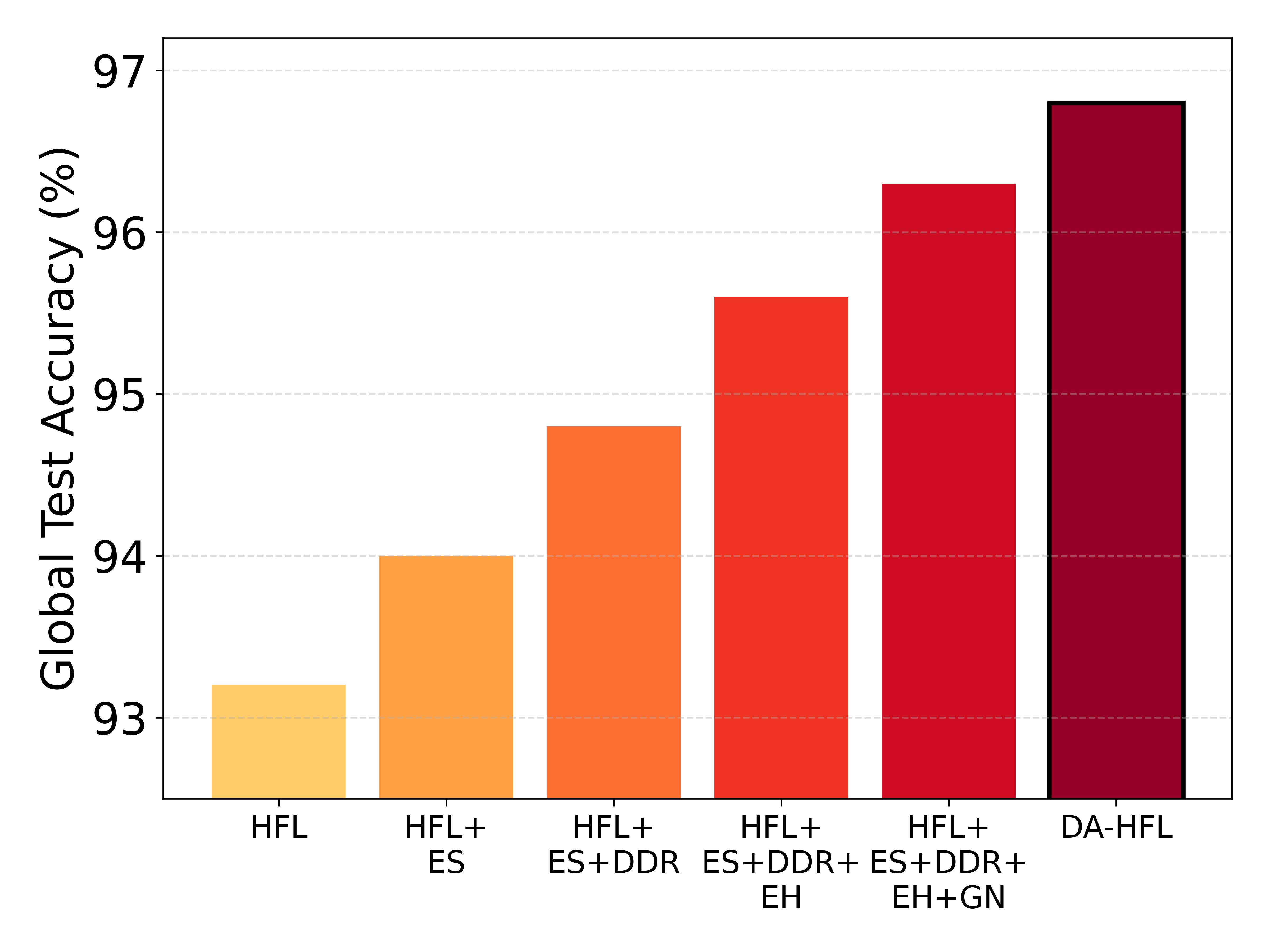}
}
\caption{(a) Accuracy across environmental zones,
(b) accuracy under domain drift, (c) cluster purity,
(d) cluster lifetime, (e) energy consumption, and
(f) ablation study.}
\label{fig:performance_comparison_uav}
\end{figure}
Fig.~\ref{fig:convergence}(a)--(b) presents test
accuracy and training loss over communication
rounds. All methods improve with training, but
\ac{DA-HFL} achieves higher accuracy and lower
loss after the common initial point. Its accuracy
reaches approximately 90\% by round 50 and
97\% by round 100. The separation between the
curves indicates that its advantage develops
during training and persists at the final round.
For heterogeneous environments, a shared
representation can retain common information,
while separate prediction heads accommodate
environment-specific class relationships.
The observed behavior is consistent with this
design, although the global curves do not
isolate the effect of either component.
Fig.~\ref{fig:convergence}(c) summarizes accuracy
over the last five rounds. The smaller error
bars indicate lower temporal variation during
this interval rather than variability across
independent experimental runs. Fig.~\ref{fig:performance_comparison} compares precision,
F1-score, recall, and ROC-AUC.
\Ac{DA-HFL} achieves approximately 97.1\%,
96.8\%, 96.5\%, and 98.8\%, respectively,
exceeding the compared methods across all
four metrics. The precision and recall results
show that the accuracy improvement is accompanied
by a favorable balance between false-positive
and false-negative predictions. The higher
F1-score summarizes this balance, while the
ROC-AUC result indicates stronger discrimination
across classification thresholds. Thus, the
reported improvement extends beyond accuracy
alone.

Fig.~\ref{fig:matrices} examines class confusion
within each environment. In city, the
vehicle-to-building entry decreases from 29
for \ac{HFL} to 18 for \ac{DA-HFL}, while
the reverse entry decreases from 21 to 13.
In forest, person-to-vegetation decreases
from 36 to 17, and water-to-person in lake
decreases from 21 to 14. These are among the
class pairs affected by the constructed
environmental ambiguity. Their lower confusion
entries are consistent with the intended role
of environment-specific heads in retaining
local class distinctions. The remaining errors
show that specialization does not completely
resolve overlapping class representations.
These values are treated as displayed matrix
entries rather than normalized percentages.

Fig.~\ref{fig:performance_comparison_uav}(a) compares accuracy across the four environments. \Ac{DA-HFL} achieves 95.6\% in city, 96.2\% in village, 97.1\% in forest, and 96.5\% in lake, showing that the aggregate gain is not confined to a single zone despite differing class ambiguity. Fig.~\ref{fig:performance_comparison_uav}(b) evaluates the forest-to-city transition at mission-time coordinate 5: \Ac{DA-HFL} retains approximately 96\% accuracy at the transition and approaches 97\%, while final accuracies are 82\% (FedAvg), 87\% (AsyncFL), 90\% (CFL), 93\% (\ac{HFL}), and 95\% (RA-HFL). Signature-based head selection, prediction mixing, and local normalization adaptation provide a rationale for this transition robustness, although the curve reflects their combined rather than individual effects.

Fig.~\ref{fig:performance_comparison_uav}(c)--(d)
reports cluster purity and lifetime.
\Ac{DA-HFL} achieves 97.3\% purity and a
35.2-round lifetime, compared with 94.2\%
and 30.6 rounds for RA-HFL.
The improvements are 3.1 percentage points
and 4.6 rounds, respectively.
Higher purity indicates stronger environmental
agreement among clustered members under the
reported measure, while a longer lifetime
indicates more persistent cluster organization.
These observations are consistent with using
environment signatures to guide membership
and transition risk to guide \ac{CH} selection.

Fig.~\ref{fig:performance_comparison_uav}(e)
compares modeled energy consumption.
\Ac{DA-HFL} consumes 8.9~J, compared with
8.5~J for RA-HFL, 12.6~J for FedAvg,
and 11.5~J for AsyncFL. Its reported cost is therefore 0.4~J higher
than RA-HFL but lower than the two flat
baselines. Together with the accuracy and
clustering results, this indicates a tradeoff
between learning performance and modeled
resource consumption. The comparison includes
the modeled training, transmission, and
propulsion costs; it does not quantify the
energy required for gated inference.
Finally, Fig.~\ref{fig:performance_comparison_uav}(f)
examines cumulative component additions.
Accuracy increases from 93.2\% for \ac{HFL}
to 94.0\% after adding environment similarity
(ES), 94.8\% with domain-drift risk (DDR),
95.6\% with environment heads (EH), and
96.3\% with the gate (GN). The complete
framework achieves 96.8\%.
The progression supports the benefit of
combining environment-aware clustering,
specialized prediction, and adaptive inference.
Because each component is added to the
preceding configuration, the observed gains
describe cumulative improvements rather than
independent component effects.

\section{Conclusion}
This paper presented \ac{DA-HFL} for mobile \ac{UAV} swarms, combining drift-aware clustering, a shared backbone with environment-specific heads, Fisher-weighted cloud aggregation, and gated inference with local adaptation. The analysis established conditional stationarity, the quadratic optimality of Fisher-weighted merging, and complexity bounds. Simulations showed faster learning, higher accuracy during environmental transitions, and purer, longer-lived clusters than the evaluated baselines, with modest energy overhead relative to resource-aware \ac{HFL}. Future work will consider learned environment boundaries, unseen-environment discovery, and physical \ac{UAV} validation.

\bibliographystyle{IEEEtran}
\bibliography{ref}

@article{saad2020vision,
  author  = {Walid Saad and Mehdi Bennis and Mingzhe Chen},
  title   = {A Vision of {6G} Wireless Systems: Applications,
             Trends, Technologies, and Open Research Problems},
  journal = {IEEE Network},
  year    = {2020},
  volume  = {34},
  number  = {3},
  pages   = {134--142},
  doi     = {10.1109/MNET.001.1900287}
}

@inproceedings{mcmahan2017communication,
  author    = {McMahan, H. Brendan and others},
  title     = {Communication-Efficient Learning of Deep Networks from Decentralized Data},
  booktitle = {Proceedings of the 20th International Conference on Artificial Intelligence and Statistics},
  series    = {Proceedings of Machine Learning Research},
  volume    = {54},
  pages     = {1273--1282},
  year      = {2017},
}

@inproceedings{li2020federated,
  author    = {Li, Tian and others},
  title     = {Federated Optimization in Heterogeneous Networks},
  booktitle = {Proceedings of Machine Learning and Systems (MLSys)},
  volume    = {2},
  pages     = {429--450},
  year      = {2020}
}

@inproceedings{karimireddy2020scaffold,
  author    = {Karimireddy, Sai Praneeth and others},
  title     = {{SCAFFOLD}: Stochastic Controlled Averaging for Federated Learning},
  booktitle = {Proceedings of the 37th International Conference on Machine Learning (ICML)},
  series    = {Proceedings of Machine Learning Research},
  volume    = {119},
  pages     = {5132--5143},
  year      = {2020},
  publisher = {PMLR}
}

@ARTICLE{10065509,
  author={Zeng, Shenglai and Li, Zonghang and Yu, Hongfang and Zhang, Zhihao and Luo, Long and Li, Bo and Niyato, Dusit},
  journal={IEEE Transactions on Cloud Computing}, 
  title={HFedMS: Heterogeneous Federated Learning With Memorable Data Semantics in Industrial Metaverse}, 
  year={2023},
  volume={11},
  number={3},
  pages={3055-3069},
  doi={10.1109/TCC.2023.3254587}}

@ARTICLE{10452821,
  author={Kang, Yufei and Li, Baochun},
  journal={IEEE Transactions on Cloud Computing}, 
  title={Polaris: Accelerating Asynchronous Federated Learning With Client Selection}, 
  year={2024},
  volume={12},
  number={2},
  pages={446-458},
  doi={10.1109/TCC.2024.3370688}}

@article{wu2020safa,
  author  = {Wu, Wentai and others},
  title   = {{SAFA}: A Semi-Asynchronous Protocol for Fast Federated Learning With Low Overhead},
  journal = {IEEE Transactions on Computers},
  volume  = {70},
  number  = {5},
  pages   = {655--668},
  year    = {2021}
}

@inproceedings{liu2020client,
  author    = {Liu, Lumin and others},
  title     = {Client-Edge-Cloud Hierarchical Federated Learning},
  booktitle = {Proceedings of the IEEE International Conference on Communications (ICC)},
  pages     = {1--6},
  year      = {2020},
  organization = {IEEE}
}

@article{wang2024uav,
  author  = {Wang, Tianshun and others},
  title   = {{UAV} Swarm-Assisted Two-Tier Hierarchical Federated Learning},
  journal = {IEEE Transactions on Network Science and Engineering},
  volume  = {11},
  number  = {5},
  pages   = {4295--4310},
  year    = {2024}
}

@article{he2023threestage,
  author  = {He, Wenji and others},
  title   = {Three-Stage Stackelberg Game Enabled Clustered Federated Learning in Heterogeneous {UAV} Swarms},
  journal = {IEEE Transactions on Vehicular Technology},
  volume  = {72},
  number  = {10},
  pages   = {13309--13322},
  year    = {2023}
}

@article{shen2023joint,
  author  = {Shen, Yun and Shen, Fangqin and Wang, Wenchao and others},
  title   = {Joint Training and Resource Allocation Optimization for Federated Learning in {UAV} Swarm},
  journal = {IEEE Internet of Things Journal},
  volume  = {10},
  number  = {2},
  pages   = {1114--1128},
  year    = {2023}
}

@article{sattler2021clustered,
  author  = {Sattler, Felix and others},
  title   = {Clustered Federated Learning: Model-Agnostic Distributed Multitask Optimization Under Privacy Constraints},
  journal = {IEEE Transactions on Neural Networks and Learning Systems},
  volume  = {32},
  number  = {8},
  pages   = {3710--3722},
  year    = {2021}
}

@inproceedings{li2021fedbn,
  author    = {Li, Xiaoxiao and others},
  title     = {{FedBN}: Federated Learning on Non-{IID} Features via Local Batch Normalization},
  booktitle = {Proceedings of the 9th International Conference on Learning Representations (ICLR)},
  year      = {2021}
}

@article{ganin2016domain,
  author  = {Ganin, Yaroslav and others},
  title   = {Domain-Adversarial Training of Neural Networks},
  journal = {Journal of Machine Learning Research},
  volume  = {17},
  number  = {59},
  pages   = {1--35},
  year    = {2016}
}

@inproceedings{li2021ditto,
  author    = {Li, Tian and Hu, Shengyuan and Beirami, Ahmad and Smith, Virginia},
  title     = {Ditto: Fair and Robust Federated Learning Through Personalization},
  booktitle = {Proceedings of the 38th International Conference on Machine Learning (ICML)},
  series    = {Proceedings of Machine Learning Research},
  volume    = {139},
  pages     = {6357--6368},
  year      = {2021},
  publisher = {PMLR}
}

@inproceedings{fallah2020personalized,
  author    = {Fallah, Alireza and Mokhtari, Aryan and Ozdaglar, Asuman},
  title     = {Personalized Federated Learning: A Meta-Learning Approach},
  booktitle = {Advances in Neural Information Processing Systems (NeurIPS)},
  volume    = {33},
  pages     = {3557--3568},
  year      = {2020}
}

@ARTICLE{11303974,
  author={Li, Jialong and Tripathi, Shreyansh and Rastogi, Lakshay and Lei, Yiming and Pan, Rui and Xia, Yiting},
  journal={IEEE Transactions on Networking}, 
  title={Optimizing Mixture-of-Experts Inference Time via Model Deployment and Communication Scheduling}, 
  year={2026},
  volume={34},
  number={},
  pages={2478-2497},
  doi={10.1109/TON.2025.3645806}}

@article{yi2026pfedmoe,
  author  = {Yi, Liping and Yu, Han and Wang, Gang and Liu, Xiaoguang and Hu, Qinghua},
  title   = {{pFedMoE}: Data-Level Personalization with Mixture of Experts for Model-Heterogeneous Personalized Federated Learning},
  journal = {IEEE Transactions on Knowledge and Data Engineering},
  volume  = {38},
  number  = {3},
  pages   = {1905--1918},
  year    = {2026},
  doi     = {10.1109/TKDE.2026.3656194}
}

@inproceedings{matena2022merging,
  author    = {Matena, Michael S. and Raffel, Colin},
  title     = {Merging Models with Fisher-Weighted Averaging},
  booktitle = {Advances in Neural Information Processing Systems (NeurIPS)},
  volume    = {35},
  pages     = {17703--17716},
  year      = {2022}
}

@article{kirkpatrick2017overcoming,
  author  = {Kirkpatrick, James and others},
  title   = {Overcoming Catastrophic Forgetting in Neural Networks},
  journal = {Proceedings of the National Academy of Sciences},
  volume  = {114},
  number  = {13},
  pages   = {3521--3526},
  year    = {2017}
}

@article{wang2022generalizing,
  author  = {Wang, Jindong and others},
  title   = {Generalizing to Unseen Domains: A Survey on Domain Generalization},
  journal = {IEEE Transactions on Knowledge and Data Engineering},
  volume  = {35},
  number  = {8},
  pages   = {8052--8072},
  year    = {2023}
}

@article{zeng2019energy,
  author  = {Zeng, Yong and Xu, Xiaoli and Zhang, Rui},
  title   = {Energy Minimization for Wireless Communication With Rotary-Wing {UAV}},
  journal = {IEEE Transactions on Wireless Communications},
  volume  = {18},
  number  = {4},
  pages   = {2329--2345},
  year    = {2019}
}

@article{dai2022energyefficient,
  author  = {Dai, Xin-Ke and others},
  title   = {Energy-Efficient {UAV} Communications: A Generalized Propulsion Energy Consumption Model},
  journal = {IEEE Wireless Communications Letters},
  volume  = {11},
  number  = {10},
  pages   = {2100--2104},
  year    = {2022}
}

@article{wang2020task,
  author  = {Guangying Wang and Xiangbin Yu and Fangcheng Xu and Jiali Cai},
  title   = {Task Offloading and Resource Allocation for UAV-Assisted
             Mobile Edge Computing with Imperfect Channel Estimation
             over Rician Fading Channels},
  journal = {EURASIP Journal on Wireless Communications and Networking},
  volume  = {2020},
  number  = {1},
  pages   = {169},
  year    = {2020},
  doi     = {10.1186/s13638-020-01780-8}
}

\end{document}